\documentclass[twocolumn,showpacs,preprintnumbers,amsmath,amssymb]{revtex4}
\usepackage{graphicx}
\usepackage{dcolumn}
\usepackage{bm}

\begin{document}

\preprint{Ravindran et al }

\title{Origin of magnetoelectric behavior in BiFeO$_{3}$}
\author{P. Ravindran}
\email{ravindran.ponniah@kjemi.uio.no}
\homepage{http://www.folk.uio.no/ravi}
\author{R. Vidya}
\author{A. Kjekshus}
\author{H. Fjellv{\aa}g}
\affiliation{Department of Chemistry,
             University of Oslo, Box 1033, Blindern N-0315, Oslo, Norway}
             \author{O. Eriksson}
\affiliation{Department of Physics,
             Uppsala University, P.O. Box 530, S-751 21, Uppsala, Sweden}
\date{\today}
\begin{abstract}
The magnetoelectric behavior of BiFeO$_3$ has been explored on the
basis of accurate density functional calculations. We are able to
predict structural, electronic, magnetic, and ferroelectric
properties of BiFeO$_3$ correctly without including any strong
correlation effect in the calculation. Unlike earlier
calculations, the equilibrium structural parameters are found to
be in very good agreement with the experimental findings. In
particular, the present calculation correctly reproduced
experimentally-observed elongation of cubic perovskite-like
lattice along the [111] direction. At high pressure we predicted a
pressure-induced structural transition from rhombohedral ($R3c$)
to an orthorhombic ($Pnma$) structure. The total energy
calculations at expanded lattice show two lower energy
ferroelectric phases (with monoclinic $Cm$ and tetragonal $P4mm$
structures), closer in energy to the ground state phase.
Spin-polarized band-structure calculations show that BiFeO$_3$
will be an insulator in A- and G-type antiferromagnetic phases and
a metal in C-type antiferromagnetic, ferromagnetic configurations,
and in the nonmagnetic state.
Chemical bonding in BiFeO$_3$ has been analyzed using
partial density-of-states, charge density, charge transfer,
electron localization function, Born-effective-charge tensor, and
crystal orbital Hamiltonian population analyses. Our electron
localization function analysis shows that stereochemically active
lone-pair electrons are present at the Bi sites which are
responsible for displacements of the Bi atoms from the
centro-symmetric to the noncentrosymmetric structure and hence the
ferroelectricity.
  A large ferroelectric polarization of
88.7\,$\mu$C/cm$^{2}$ is predicted in accordance with recent
experimental findings, but differing by an order of magnitude from
earlier experimental values. The strong spontaneous polarization
is related to the large values of the Born-effective-charges at
the Bi sites along with their large displacement along the [111]
direction of the cubic perovskite-type reference structure. Our
polarization analysis shows that partial contributions to
polarization from the Fe and O atoms almost cancel each other and
the net polarization present in BiFeO$_3$  mainly ($>$ 98\%)
originates from Bi atoms. We found that the large scatter in
experimentally reported polarization values in BiFeO$_3$ is
associated with the large anisotropy in the spontaneous
polarization.
\end{abstract}
\maketitle

\section{INTRODUCTION}
\label{sec:intro} Multiferroic materials have coupled electric,
magnetic and/or structural order parameters that result in
simultaneous ferromagnetic, ferroelectric and/or ferroelastic
behavior. The magnetoelectric materials have two order parameters,
viz. spontaneous polarization for ferroelectric or
antiferroelectric cases and spontaneous magnetization for
ferromagnetic, ferrimagnetic  or antiferromagnetic cases. Coupling
between the spontaneous polarization and spontaneous magnetization
lead to magnetoelectric (ME) effects, in which the magnetization
can be tuned by an applied electric field and vice versa. There
has been considerable recent
interest~\cite{fiebig02,kimura03,fiebig04} in developing
multiferroic materials in which interaction between the magnetic
and electronic degrees of freedom allows both charge and spin to
be manipulated by applied electric/magnetic
field~\cite{smolenskii82,schmid94}. These materials have
opportunities for potential applications in information storage,
the emerging field of spintronics, and sensors. Other potential
applications include multiple state memory elements, in which data
is stored both in the electric and the magnetic polarizations, or
novel memory media which might allow writing of a ferroelectric
data bit, and reading of the magnetic field generated by
association. Besides the potential applications, the fundamental
physics of magnetoelectric materials is rich and fascinating.
Hence, the theoretical understanding of the intrinsic physical
properties of multiferroic oxides is clearly of great importance
for both fundamental science and technological applications.

\par
BiFeO$_3$ is ferroelectric with relatively high Curie temperature
($T_{\rm C}; ca. $ 1100 K) and exhibits antiferromagnetic behavior
with high N\'{e}el temperature ($T_{\rm N}; ca. $ 643
K)~\cite{bhide65,moreau71,venevtsev95}. The Fe magnetic moments
are coupled ferromagnetically (F) within the pseudocubic (111)
planes and antiferromagnetically (AF) between adjacent planes. If
the magnetic moments are oriented perpendicular to the [111]
direction, the symmetry also permits a canting of the AF
sublattices resulting in a macroscopic magnetization; so-called
weak ferromagnetism~\cite{dzyaloshinskii57,moriya60}. The magnetic
structure of BiFeO$_3$ is, in the first approximation, AF with
G-type magnetic ordering (G-AF). But the G-AF has been modified by
subjecting it to a long range-modulation, manifesting itself in a
cycloidal spiral of $\lambda$= 620\,{\AA} length with [110] as the
spiral propagation direction and spin rotation within (110) which
is unusual for perovskites~\cite{sosnowska82}. BiFeO$_3$ shows the
linear magnetoelectric effect, with applied magnetic fields
inducing weak F and large increase in polarization~\cite{popov93}.
The exhibition of weak magnetism at room temperature is due to a
residual moment from the canted spin
structure~\cite{smolenskii82}. A significant magnetization ($\sim
1\mu_{B}$ per unit cell), as well as a strong ME coupling, have
been reported recently in high quality thin films~\cite{wang03}.
Coexistence of ferroelectricity and ferromagnetism has been
observed~\cite{kim03} in bulk BiFeO$_{3}$-PrFeO$_{3}$-PbTiO$_{3}$
solid solutions with compositions
0.2BiFeO$_{3}$-0.2PrFeO$_{3}$-0.6PbTiO$_{3}$ and
0.4BiFeO$_{3}$-0.2PrFeO$_{3}$-0.6PbTiO$_{3}$. Significant
dielectric enhancement in ferromagnetic
0.3BiFeO$_{3}$-0.7SrBi$_{2}$Nb$_{2}$O$_{9}$ has been
reported~\cite{gu01}, for samples synthesized by sintering
mechanically activated oxide mixture. Since epitaxial films of
BiFeO$_3$ also show large electric polarization ($\sim 158 \mu$
C/cm$^{2}$)\cite{yun04} they are promising candidate materials for
ME-device applications.

\par
An important aspect that emerges upon examination\cite{teague70}
of the properties of single crystal BiFeO$_3$ is that it has a
spontaneous polarization that is significantly smaller than the
expected value for a ferroelectric with such a high T$_C$. It was
until recently not clear  whether this is a result of intrinsic
material properties or of limitations imposed by leakage and
imperfect material quality in bulk. Enhancement of spontaneous
polarization and related properties have been
reported~\cite{wang03,li04,yun04} for heteroepitaxially
constrained thin films of BiFeO$_3$  recently and  more recent
theoretical calculations~\cite{neaton05,ederer051} also predicated
correctly the presence of large polarization in this material. The
sizable polarizations in BiFeO$_3$ are consistent with the
observed large atomic distortions~\cite{michel69,kubel90}, but
apparently inconsistent with earlier studies of bulk
BiFeO$_3$~\cite{teague70,palkar02,wang04}, origin of these
distinction being currently under debate. This is one of the
motivations for the present study.
\par
Recent theoretical report on structural behaviors of BiFeO$_3$
using density functional theory (DFT) calculations show
considerable deviation from the experimental
findings~\cite{neaton05,wang03}. An ideal cubic perovskite-type
structure can alternatively be described in a rhombohedral cell
with rhombohedral angle ($\alpha_{R}$)= 60$^o$. The low
temperature neutron measurements~\cite{sosnowska02} show
$\alpha_{R}$ = 59.344$^{\textnormal{o}}$ and
59.35$^{\textnormal{o}}$ from single crystal X-ray diffraction
measurements~\cite{kubel90}. This can be rationalized as a
elongation of an ideal cubic perovskite-type structure along the
[111] direction. In contrast, the hitherto available theoretical
results available~\cite{neaton05,wang03} for BiFeO$_3$ conclude
with  $\alpha_{R}$ = 60.36$^{\textnormal{o}}$  which requires
compression  along the [111] direction of the cubic perovskite
like lattice, just opposite to the experimental findings. In order
to find out whether DFT has failed to reproduce the structural
parameters correctly for BiFeO$_3$, accurate calculations are
needed. This is another motivation for pursuing this work.

\par
Moreover, observation of enhanced polarization in
heteroepitaxially constrained thin films of BiFeO$_3$, and the
structural analysis~\cite{wang03} suggest an approximately
tetragonal structural arrangement with a superimposed  monoclinic
distortion of about 0.5$^{o}$. So, it is indeed interesting to
look for possible metastable phases with slightly distorted
structural arrangements to identify stable structure of the
experimentally-obtained heteroepitaxially-constrained thin films.
Also, structural stability studies on BiFeO$_3$ would be useful to
identify the paraelectric phase. Total energy studies on
ferroelectric BiFeO$_3$ in different magnetic configurations would
give more insight into its magnetic properties. Hence, we have
made full structural optimization for BiFeO$_3$ in different
atomic arrangements also including different magnetic
configurations in the calculations.

\par
The ``lone pair" in Bi or Pb-based oxides is believed to form due
to the hybridization of 6$s$ and 6$p$ atomic orbitals with
6$s^{2}$ electrons filling one of the resulting orbitals in Bi or
Pb in their oxides. The lone pair is then considered to be
chemically inactive, not taking part in the formation of bonds but
sterically active. The hybridization causes the lone pair to lose
its spherical symmetry and is projected out on one side of the
cation, resulting in an asymmetry of the metal coordination and
distorted crystal structures. In the present study we have
analyzed the occurrence of lone pair and its role on structural
stability with the help of charge density, charge transfer,
electron localization function, crystal orbital Hamiltonian
population, and partial density of states analyses.

\par
The long-range Coulomb interaction plays a crucial role in
ferroelectric materials. It is therefore critical to be able to
calculate the spontaneous (i.e., zero field) electric polarization
{\bf P} as a function of structural degrees of freedom. It has
been shown~\cite{kingsmith93,resta92} that {\bf P} can directly be
calculated as a Berry phase of the Bloch states. As for the
dependence of {\bf P} upon the structure, the first-order
variation of {\bf P} with atomic displacements is given by the
dynamic effective charge $Z^{*}$, which can be computed either
directly using linear-response methods~\cite{baroni87} or
numerically by finite differences. For the present study Born
effective charges were obtained from finite differences of
macroscopic polarization induced by small displacements of the
atomic sublattices. Recent theoretical
reports~\cite{neaton05,ederer05} indicated that it is important to
include Coulomb correlation effects via the LDA+$U$ approach into
the calculation to describe the electronic structure and
ferroelectric behavior of BiFeO$_3$. This motivated us to perform
accurate DFT calculation  to check the possibility of describing
electronic structure, magnetism, and ferroelectric behavior of
BiFeO$_3$ within the usual DFT framework.

\par
In this paper we report results of first principle DFT
calculations on structural optimization for BiFeO$_3$ in different
atomic arrangements and also with different magnetic
configurations. Moreover, we investigate structural phase
stability, the electronic structure, chemical bonding, magnetism,
and ferroelectric properties of BiFeO$_3$. The remainder of this
paper is organized as follows. In section \ref{sec:compuational},
we describe the structural aspects and the computations involved
in the evaluation of structural phase stability, magnetism, and
ferroelectric behaviors. In Sec.\,\ref{sec:resdis} we present our
results on structural phase stability, electronic structure,
chemical bonding, magnetism, and ferroelectricity, and compare the
findings with available experimental and theoretical studies. We
summarize and conclude in Sec.\,\ref{sec:con}.

\section{Computational details}
\label{sec:compuational} First principles DFT calculations were
performed using the Vienna {\em abinitio} simulation package
(VASP)~\cite{vasp} within the projector augmented wave (PAW)
method~\cite{paw} as implemented by Kresse and
Joubert~\cite{kresse99}. The Kohn-Sham equations~\cite{kohn65}
were solved self-consistently using an iterative matrix
diagonalization method. This is based on a band-by-band
preconditioned conjugate gradient~\cite{payne92} method with an
improved Pulay mixing~\cite{pulay80} to efficiently obtain the
ground-state electronic structure. The forces on the atoms were
calculated using the Hellmann-Feynman theorem and they are used to
perform a conjugate gradient relaxation. Structural optimizations
were continued until the forces on the atoms had converged to less
than 1~meV/{\AA} and the pressure on the cell had minimized within
the constraint of constant volume. The calculations were performed
within periodic boundary conditions allowing the expansion of the
crystal wave functions in terms of a plane-wave basis set.

\par
 The Generalized Gradient Approximation
(GGA)~\cite{perdew} includes the effects of local gradients in the
charge density for each point  which generally gives better
equilibrium structural parameters than the local density
approximation (LDA). Ferroelectric properties are extremely
sensitive to structural parameters (lattice parameters and atom
positions) and hence we have used GGA for all our studies.  In the
basis we treated explicitly 15 valence electrons for Bi
(5$d^{10}$6$s^{2}$6$p^{3}$), 14 for Fe (3$p^{6}$3$d^{6}$4$s^{2}$,
and 6 for oxygen (2$s^{2}$2$p^{4}$). Brillouin zone integrations
are performed with a Gaussian broadening~\cite{elsasser94} of
0.1\,eV during all structural optimizations. These calculations
are performed with a 4$\times$4$\times$4 Monkhorst-Pack {\bf
k}-point mesh\cite{monkhorst} centered at $\Gamma$ for the
ferroelectric $R3c$ structure and similar {\bf k}-point density
has been used for all the other structures considered for the
present study. One specific problem of these materials is the
presence of ``computationally-unfriendly" atom configurations such
as the first-row element oxygen and  the transition metal iron.
Both are kinds of atoms which require extra care: either large
basis sets within a pseudopotential scheme, or an all-electron
scheme. So we have used very large basis set with 875~eV for the
plane-wave cutoff. For the $\bf{k}$-space integrations in the
Berry-phase calculations, a uniform 8 $\times$ 8 $\times$ 8
$\bf{k}$-point mesh was found to be adequate.
\par
BiFeO$_3$ has a rhombohedrally distorted perovskite-type structure
with space group $R3c$~\cite{michel69,kubel90}. For the study of
structural phase stability we have considered eight closely
related potential structure ``types" : BiFeO$_3$ (rhombohedral;
$R3c$)~\cite{kubel90}, CaTiO$_3$ (perovskite prototype;
cubic;$Fm$\={3}$m$)~\cite{catio3}, KNbO$_3$ (tetragonal;
$P4mm$)~\cite{knbo3}, $\beta$-LaCoO$_{3}$ (rhombohedral;
$R$\={3}$c$)~\cite{lacoo3}, $\alpha$-LaCoO$_3$
(monoclinic;$I2/a$)~\cite{a-lacoo3}, PbFe$_{0.5}$Nb$_{0.5}$O$_3$
(monoclinic; $Cm$)~\cite{c1m1}, GdFeO$_3$ (orthorhombic $Pnma$),
and LaVO$_3$ (monoclinic; $P2_{1}/c$)~\cite{lavo3}. The reasons
for choosing these structure types for our calculations are given
below. It is experimentally reported that BiFeO$_3$ stabilizes in
the rhombohedral $R3c$ and hence it is natural to include this
structure variant. This structure which is also adopted by the
paraelectric $\beta$-LaCoO$_{3}$ structure  can be derived from
the ideal cubic perovskite-type structure by rotation of the
octahedra along the cubic [111] axis. Since the experimental
studies show enhancement of the polarization in heteroepitaxially
constrained thin films of BiFeO$_3$, for which the structural
analyses~\cite{wang03} suggest an approximately tetragonal crystal
structure with a small monoclinic distortion, we considered
structural arrangements within space groups  $P4mm$ and $Cm$.
Recent experimental studies show that the $\beta$-LaCoO$_{3}$-type
rhombohedral structure undergoes a small monoclinic distortion and
stabilizes in an $\alpha$-LaCoO3 type structure~\cite{a-lacoo3}.
The orthorhombic GdFeO$_3$-type structure arises from the ideal
cubic perovskite by tilting of the FeO$_6$ octahedra around the
cubic [110] axis and with small monoclinic distortion one arrives
at the LaVO$_3$-type structure~\cite{lavo3}. As experimentally
reported crystal structure of thin film has small monoclinic
distortion, we have considered the monoclinic I2/a, $Cm$, and
$P2_{1}/c$ structures in our investigations.

\par
The structures were fully relaxed for all volumes considered in
the present calculations using force as well as stress
minimization. Experimentally established structural data were used
as input for the calculations. In order to avoid ambiguities
regarding the free-energy results we have always used the same
energy cutoff and corresponding {\bf k}-grid densities for
convergence in all calculations. A criterion of at least
0.01~meV/atom was placed on the self-consistent convergence of the
total energy. Brillouin zone integration was performed with a
Gaussian broadening of 0.1\,eV during all relaxations. The present
type of theoretical approach has recently been successfully
applied~\cite{vajlet,vajapl} to reproduce experimentally observed
ambient- and high-pressure phases of metal hydrides.
\par
The experimental studies show that the ground state of BiFeO$_3$
is AF with a N\'{e}el temperature of
643\,K~\cite{kiselev63,teague70}. A noticeable magnetization
($\sim$1$\mu_{B}$ per cell), as well as a strong magneto-electric
coupling have been reported recently in high-quality epitaxial
thin films~\cite{wang03}. The Fe magnetic moments are coupled F
within the pseudocubic (111) planes and AF between adjacent
planes. It has been reported as a long-wavelength spiral spin
structure~\cite{sosnowska82} and possibly a small out-of-plane
canting due to weak ferromagnetism~\cite{vorobev95}.
Owing to tilting of FeO$_{6}$ octahedra, Fe-O-Fe chains are not
subtending to an angle of 180$^o$. As the exchange interaction
takes place via these Fe-O-Fe chains, canted spin arrangements
arise which further lead to antiferromagnetism with weak
ferromagnetism.  Since the noncollinearity of the magnetic moments
appears to be quite minimal, we made the computational
simplification to approximate the magnetic structure to a
collinear model.
\par
\begin{figure*}
\begin{minipage}{\textwidth}
\hspace{-0.40in}
\includegraphics[scale=0.90]{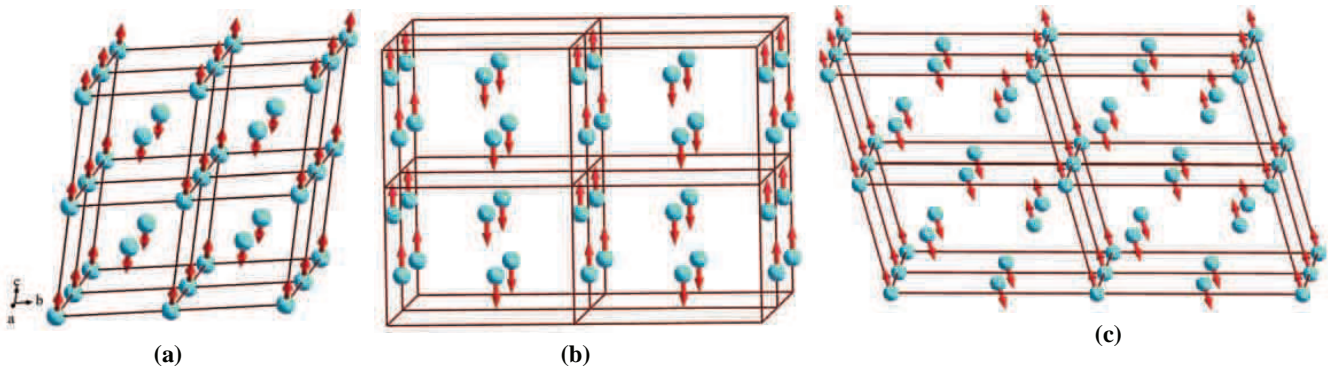}
\caption{\label{fig:magstr}(Color online) Antiferromagnetic
structures considered for ferroelectric BiFeO$_3$ in the (a)
A-AF-like structure, (b) C-AF-like structure , and (c) G-AF-like
structure. For clarity only the Fe atoms are shown in a 2 $\times$
2 $\times$ 2 of the primitive cell used in the calculation. }
\end{minipage}
\end{figure*}

BiFeO$_3$ has three possible AF arrangements depending on the
interplane and intraplane couplings within structure. (i) With
interplane AF coupling and intraplane F coupling the A-AF
structure arises. (ii) The opposite structure, with the interplane
F coupling and intraplane AF coupling is called C-AF arrangement.
In the C-AF magnetic structure all magnetic atoms have two F and
four AF nearest neighbors (vice-versa in the case of A-AF). (iii)
If both the inter- and intraplane couplings are AF the G-AF
arrangement arises. In a G-AF arrangement for an ideal cubic
perovskite-type structure, each magnetic atom is surrounded by six
AF neighbors. Because of the rhombohedral distortion we were
unable to introduce perfect G-AF ordering (see
Fig.~\ref{fig:magstr}). However, the adopted G-AF magnetic
arrangement has perfect AF ordering according to cubic
perovskite-type structure along the $ac$ plane and zig-zag chains
with AF ordering in the $bc$ plane. For the structural phase
stability calculations we assumed AF ordering with {\bf q} vector
(0,0,1/2) for all structures except for the cubic variant where we
have assumed G-AF ordering and for the R3c variant where we
considered nonmagnetic, F, A-AF, C-AF and G-AF orderings.

\par
In order to understand the role of Bi lone pair on the
ferroelectric properties of BiFeO$_3$ we have visualized it with
the help of electron localization function (ELF). The ELF provides
a measure of the local influence of the Pauli repulsion on the
behavior of the electrons and permits mapping in real space of
core, bonding, and nonbonding regions in a crystal. It can be used
to distinguish the nature of different types of bonding in
solids~\cite{savin97}. It is defined as

\begin{equation}
ELF = [1 + (\frac{D}{D_{h}})^{2}]^{-1}, \label{eqn:eq1}
\end{equation}

where \[ D = \frac{1}{2}\sum_{i}|\nabla\phi_{i}|^{2} - \frac{1}{8}
\frac{|\nabla\rho|^{2}}{\rho} \]

and \[D_{h} = \frac{3}{10}(3\pi^{2})^{5/3}\rho^{5/3} \] here
$\rho$ is the electron density and $\phi_{i}$ are the Kohn-Sham
wave functions.

\section{Results and discussions}
\label{sec:resdis}
\subsection{Structural phase stability}

\begin{figure}
\includegraphics[scale=0.5]{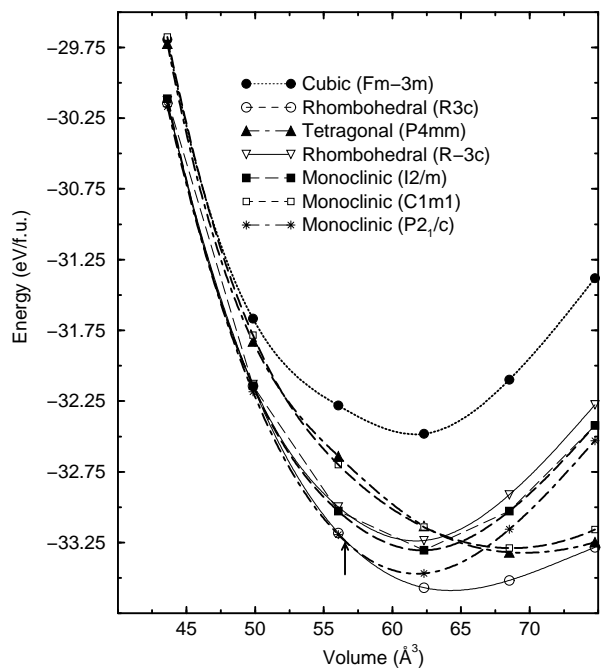}
\caption{ Calculated cell volume vs total energy for
antiferromagnetic BiFeO$_3$ in different possible structural
arrangements (structure types being labelled on the illustration).
The arrow indicates the pressure-induced structural transition
point from the rhombohedral ($R3c$) to the orthorhombic ($Pnma$)
structure.} \label{fig:allstr}
\end{figure}

BiFeO$_3$ is reported to exhibit about eight structural
transitions and a weak F ordering in its ground state
structure~\cite{krainik82,polomska74}. The structural instability
in perovskite-like oxides is partly related to the lattice
mismatch between the cations. Such a mismatch is conveniently
quantified in terms of the empirical Goldsmith tolerance factor
$t$. It is defined as the ratio between the ideal cubic lattice
parameters based on A$-$O vs B$-$O bonding alone, with $t=1$
corresponding to high stability of the cubic perovskite structure.
Using the ionic radii given by Shannon and
Prewitt~\cite{shannon69}, the calculated tolerance factor [$t =
(r_{Bi} + r_{O}/\sqrt{2}(r_{Fe}+r_O$)] for BiFeO$_3$ is 0.96.
Relatively low value of $t$ for BiFeO$_3$ suggests that Bi ion may
be unstable in the high-symmetric location possessed by the cubic
perovskite-type structure, viz. consistent with the total energy
results discussed below(shown in Fig:\,\ref{fig:allstr}). In order
to minimize the lattice mismatch and gain stability, a lattice
distortion arises by the cooperative rotation of FeO$_6$ octahedra
along the [111] direction. In the ideal cubic perovskite
structure, the BiO$_3$ planes are equidistant from the neighboring
Fe (111) planes. Due to the rotation of the octahedra, three
Bi$-$O distances are reduced, and three increased, giving rise to
an octahedral distortion with rhombohedral symmetry.

\par
\begin{figure}
\includegraphics[scale=0.75]{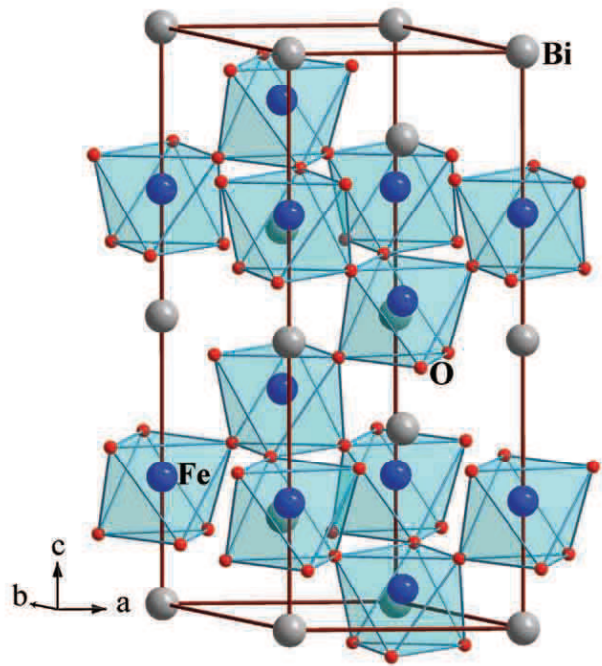}
\caption{(Color online) Crystal structure of BiFeO$_3$ in the
ferroelectric $R3c$ structures. Highly distorted FeO$_6$ octahedra
are corner shared through oxygen. Also note that Bi atoms are not
present exactly in the middle between the FeO$_6$ octahedra owing
to the off-center displacement of Bi due to the presence of 6$s$
lone pair electrons.} \label{fig:str}
\end{figure}

The structure of the ferroelectric BiFeO$_3$ phase has been
resolved experimentally using x-ray and neutron
diffraction~\cite{fischer80,kiselev63,michel69,bucci72,teague70,roginskaya66,kubel90},
and found to possess a highly distorted perovskite structure with
rhombohedral symmetry and space group $R3c$. The unit cell of
BiFeO$_3$ arising from the cubic perovskite structure by two kinds
of distortions is shown in Fig.\,\ref{fig:str}.  One distortion is
the polar displacements of all the anion and cation sublattices
relative to each other, which lead to the spontaneous electric
polarization, and the other is an antiferro-distortive rotation of
the FeO$_6$ octahedra along the [111] direction with alternating
sense of rotation along the [111] axis (owing to the lattice
mismatch as mentioned above). In terms of symmetry groups, the
polar displacements alone would reduce the symmetry from cubic
$Pm$\={3}$m$ to rhombohedral $R3m$, whereas the rotation of the
FeO$_6$ octahedra alone would lead to the paraelectric phase with
the space group $R$\={3}$c$. The incorporation of both kinds of
distortions gives the actual ferroelectric phase of BiFeO$_3$ with
the space group $R3c$.

\par
The calculated total energy vs volume for BiFeO$_3$ in fully
relaxed potential structures in the AF states are given in
Fig.\,\ref{fig:allstr}. This figure shows that the experimentally
observed ferroelectric $R3c$ phase represents the ground state
structure for BiFeO$_3$. The calculated equilibrium structural
parameters for BiFeO$_3$ in the ground state and closely related
structures are given in Table\,\ref{table:str}. Our calculated
equilibrium structural parameters are also compared with available
experimental and theoretical values in this table. The equilibrium
volume obtained from the calculation is overestimated compared to
the experimental volume by $\sim$ 3\% and the lattice parameters
are overestimated by only 1.1\%. The overestimation of equilibrium
volume by GGA for rhombohderal phase of ferroelectric materials
has already been observed for BaTiO3 and PbTiO3~\cite{wu04}, small
deviation between experimental and theoretical values indicate the
predictive capabilities of the present type of state-of-the-art
density functional calculations. It should be noted that the same
method with lower energy cut-off and {\bf k}-point densities
yielded~\cite{neaton05} the equilibrium volume which is much
smaller ($\sim$ 6.9\%) than the experimental values. An earlier
report~\cite{wang03} shows that the ferroelectric properties of
BiFeO$_3$ are very sensitive to the small changes in the lattice
parameters. So, accurate calculations for the estimation of
structural parameters for BiFeO$_3$ are needed in order to predict
the ferroelectric properties correctly.
\par
The hypothetical nonferroelectric cubic perovskite-type structure
is found to exhibit the highest energy among the various BiFeO$_3$
phases tested in this study. The energy difference between the
prototypical cubic structure and the equilibrium ferroelectric
$R3c$ structure is around 1.1~eV/f.u. This indicates that the
paraelectric phase of BiFeO$_3$ does not take the  cubic
perovskite-like structure. In fact several low symmetric atomic
arrangements based on tetragonal, rhombohedral, and monoclinic
structures are found at lower total energy than the cubic
perovskite-type structure (see Fig.\,\ref{fig:allstr}). It is
worthy to note that several low symmetry structures are found to
be stable at high pressures in the closely related system
PbTiO$_3$~\cite{wu05}. The stabilization of low symmetric
structures is associated with the presence of lone pair electrons
on Bi and also the hybridization between Bi and O. At expanded
lattice we have identified two ferroelectric structures which are
energetically closer to the ground state structure. One is a
tetragonal $P4mm$ structure and another is a monoclinic $Cm$
structure. According to a recent structural refinement by Noheda
et al.~\cite{noheda00} PbZr$_{0.52}$Ti$_{0.48}$O$_{3}$ (PZT) at
20~K is in the monoclinic $Cm$ structure.  PZT also stabilizes in
the tetragonal $P4mm$ structure. This material exhibits extremely
large electromechanical coupling in both tetragonal and monoclinic
phases. It is also interesting  to note that enhancement of
spontaneous polarization and related properties in
heteroepitaxially constrained thin films of BiFeO$_3$ has been
recently reported~\cite{wang03,li04,yun04}. In this context
observation of metastable $Cm$ and $P4mm$ phases in BiFeO$_3$ is
particularly interesting.
\par
The total energy curves in Fig.\,\ref{fig:allstr} show that there
is a large difference in total energy between considered phases at
the expanded lattice. However, at high pressures the energy
difference is small which is associated with the weakening of
directional bonding at higher pressures. Interestingly  we
observed a pressure-induced structural transition from the ground
state rhombohedral R3c structure to a monoclinic P2$_1$/c
structure around 13~GPa. As the monoclinic distortion was
negligibly small, our detailed analysis of the high pressure phase
yielded an orthorhombic $Pnma$ (GdFeO$_3$-type) structure. The
predicted structural parameters for the high pressure phase at the
phase transition point are given in Table~\ref{table:str}.
\par
The calculated bulk modulus for the ground state phase and the
possible meta stable phases are also given in
Table~\ref{table:str}. The bulk moduli have been obtained using
the so-called universal equation of state fit for the total energy
as a function of volume. Experimental bulk modulus values are not
available for this compound.  The calculated bulk modulus for the
ground state and the high pressure phases are found to be almost
double the value of that for the other metastable phases given in
Table~\ref{table:str} and this is associated with the smaller
equilibrium volume of the former phases.
\begin{table*}
\caption{Optimized structural parameters and bulk modulus
($B_{0}$)  for BiFeO$_3$ in different structural arrangements.} {
\scriptsize
\begin{ruledtabular}
\begin{tabular}{l l l c c }
Modification & Unit cell  & Positional parameters & $B_0$ (GPa)   \\
(structure type)  &  dimension  (\AA) or $^{\textnormal{o}}$    & &                   \\
\hline
\\
$R3c$ (Present - theory)                       & $a$ = 5.697; $\alpha$ = 59.235 & Bi(2a): 0, 0, 0; Fe(2a): .2232, .2232, .2232 & 130.9\\
                                      &  V$_o$=128.48               & O(6b): .5342, .9357, .3865                   & \\
$R3c$ (Ref.\,\onlinecite{neaton05} - theory)   & $a$ = 5.46;  $\alpha$ = 60.36 & Bi(2a): 0, 0, 0; Fe(2a): .231, .231, .231 & --\\
                                      &  V$_o$=115.98               & O(6b): .542, .943, .408                   & \\
$R3c$ (Ref.\,\onlinecite{baettig05} - theory)   & $a$ = 5.50;  $\alpha$ = 59.99 & Bi(2a): 0, 0, 0; Fe(2a): .228, .228, .228 & --\\
                                      &  V$_o$=117.86               & O(6b): .542, .942, .368                   & \\
$R3c$ (Ref.\,\onlinecite{kubel90} - experiment)    & $a$ = 5.63;  $\alpha$ = 59.35 & Bi(2a): 0, 0, 0; Fe(2a): .221, .221, .221 & --\\
                                      &  V$_o$=124.60               & O(6b): .538, .933, .395                   & \\

$R$\={3}c (Present - theory)                   & $a$ = 5.513; $\alpha$ = 61.432 & Bi(2b): 1/2, 1/2, 1/2; Fe(2a): 0, 0, 0 & 89.8  \\
                                      &  V$_o$=122.29               & O(6b): .3358, .1641 1/4                  &\\
$R$\={3}c (Ref.\,\onlinecite{neaton05} - theory) & $a$ = 5.35; $\alpha$ = 61.93 & Bi(2b): 1/2, 1/2, 1/2; Fe(2a): 0, 0, 0 & --\\
                                      &  V$_o$=113.12               & O(6b): .417, ..083 1/4                  & \\
$Pnma$ (Present-theory) & $a$ = 5.5849; $b$= 7.6597; $c$=5.3497  & Bi(4c): 0.0536, 1/4, 0.98858; Fe(4b): 0, 0, 1/2 & 138.8\\
                                      &  V$_o$=113.12               & O(4c): .9750, 1/4, .4060; O(8d): .2000, .9540 .1945 & \\
$P4mm$ (Present-theory) & $a$ = 3.7859; $c$= 4.8525; & Bi(1a): 0, 0, .9491; Fe(1b): 1/2, 1/2, .51804 & 72.3\\
                                      &  V$_o$=139.10               & O(1b): 1/2, 1/2, .1428; O(2c): 0, 1/2,.6735 &  \\
$Cm$ (Present-theory) & $a$ = 5.7900; $b$= 5.6899; $c$= 4.1739; & Bi(2a): .9376, 0, .0685; Fe(2a): .5110, 0, ..4961 & 79.0 \\
       & $\beta$=91.99; V$_o$=137.42               & O(2a): .5626, 0, .9489; O(4b): .7958, .7603,.4231 &\\

\end{tabular}
\end{ruledtabular}
} \label{table:str}
\end{table*}
\par
\begin{figure}
\includegraphics[scale=0.5]{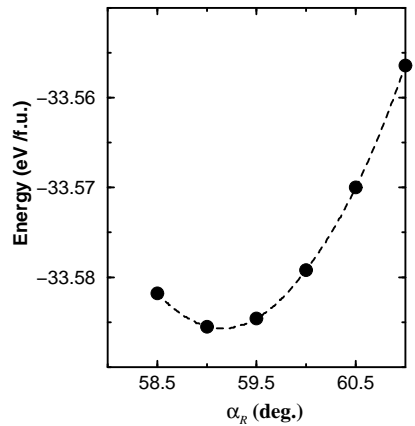}
\caption{ Total energy as a function of rhombohedral angle
($\alpha_R$) for BiFeO$_3$ in the ferroelectric $R3c$ phase.}
\label{fig:angle}
\end{figure}

The ideal cubic perovskite structure can be considered as a
special case of the rhombohedral structure with a rhombohedral
angle of 60$^{\textnormal{o}}$. The experimental
studies~\cite{kubel90,sosnowska02} show that $\alpha_{R}$ for
BiFeO$_3$ is 59.344$^{\textnormal{o}}$. In contrast, the hitherto
available theoretical results~\cite{neaton05,wang03} show that the
$\alpha_{R}$ value is 60.36$^{\textnormal{o}}$ for BiFeO$_3$ which
can only be achieved by compression of the cubic perovskite-like
lattice along the [111] direction.
With this background it is interesting to calculate the variation
in total energy as a function of the rhombohedral angle for
BiFeO$_3$ for the equilibrium volume. The calculated total energy
vs $\alpha_{R}$ given in Fig.\,\ref{fig:angle} clearly show that
the theoretical equilibrium rhombohedral angle obtained from
accurate total energy calculation is much smaller than
60$^{\textnormal{o}}$ which is consistent with experimental
observation. Compared with the total energy corresponding to the
earlier theoretically reported~\cite{neaton05,wang03} value for
$\alpha_{R}$, a gain of 12.9\,meV/f.u. is obtained (the total
energy gain from 60$^{\textnormal{o}}$ to
59.235$^{\textnormal{o}}$ is 6.5 meV/f.u.). Hence, our theoretical
study confirms the experimental observation that the rhombohedral
distortion in the ferroelectric BiFeO$_3$ phase corresponds to an
elongation rather than a compression of cubic perovskite-type
lattice along [111].

\par
We have also measured high resolution synchrotron powder
diffraction data collected at the Swiss-Norwegian beam lines at
ESRF, Grenoble, for the 2$\theta$ range from 3 $-$
38$^{\textnormal{o}}$ and $\lambda$= 0.49969 {\AA}. The structural
parameters experimentally measured at 115~K are given in
Table\,\ref{table:str} and found to be in good agreement with our
calculated equilibrium structural parameters for the ferroelectric
phase.
We also found no significant difference in the measured structural
parameters at 115~K for a sample subjected to 0.8~T magnetic field
during measurements. More details about the sample preparation and
other experimental details will be published later.

\par
\begin{figure*}
\begin{minipage}{\textwidth}
\hspace{-0.60in}
\includegraphics[scale=0.5]{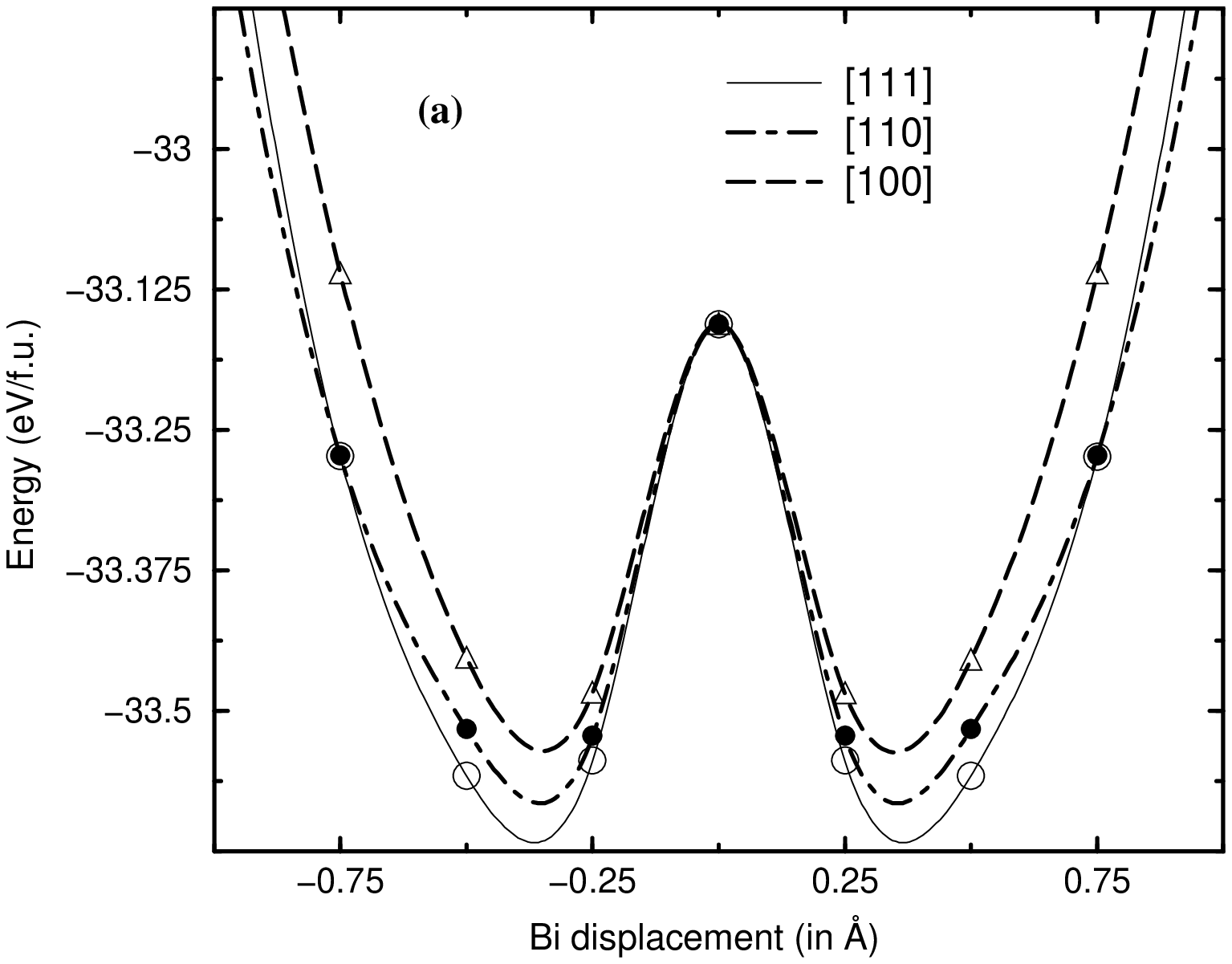}
\includegraphics[scale=0.5]{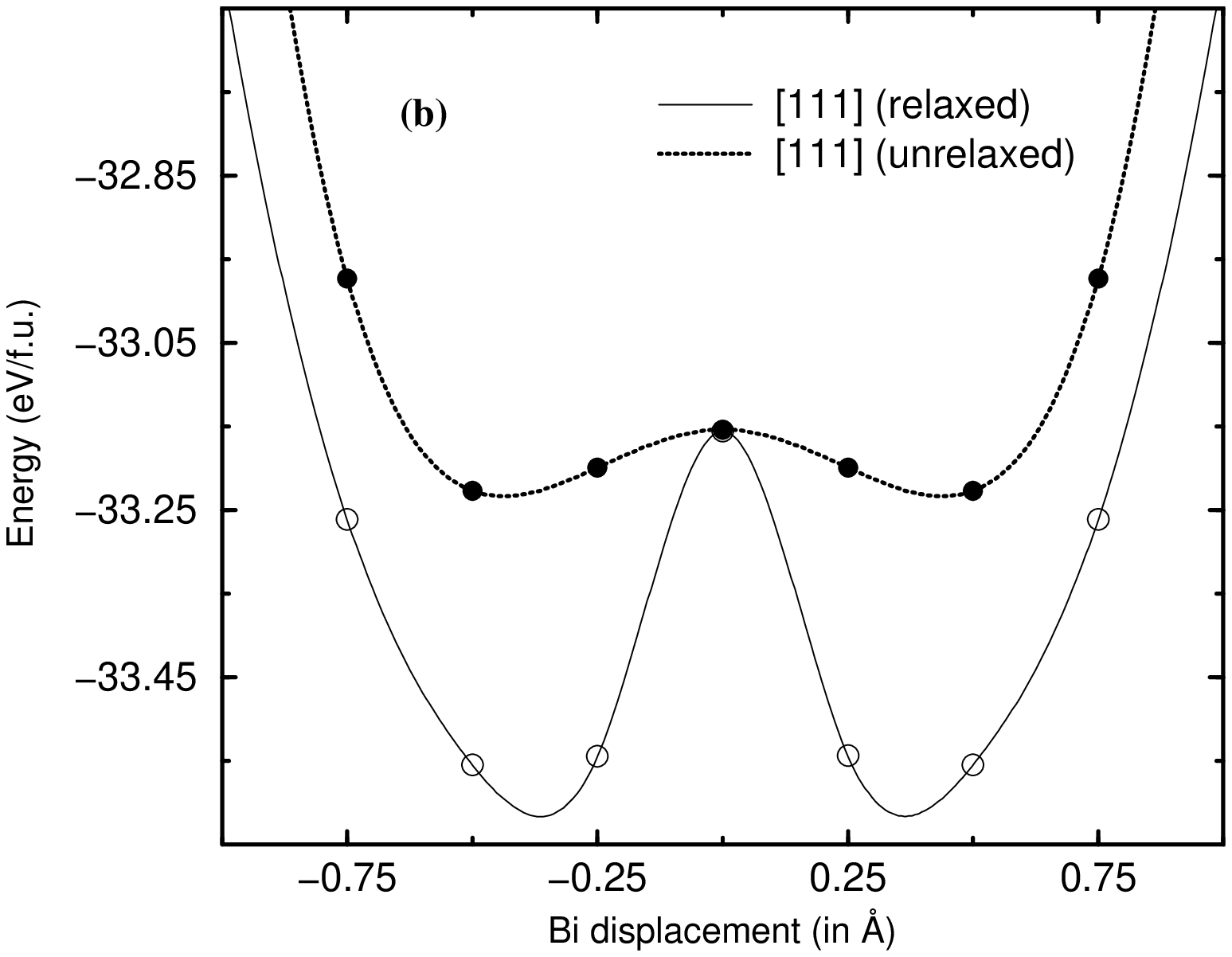}
\caption{\label{fig:dist} (a) Total energy as a function of
displacement of Bi ion  along [100], [110] and [111] directions in
the $R3c$ structure for the ferroelectric phase of BiFeO$_3$. (b)
The same for BiFeO$_3$ in the [111] direction with or without
structural relaxation.}
\end{minipage}
\end{figure*}

In order to identify the possible polarization path we mapped the
total energy as a function of displacement of Bi ions with respect
to the FeO$_6$ octahedra along [001], [110] and [111] directions
in the paraelectric R\={3}c phase in Fig.\,\ref{fig:dist}. This
helps to visualize the directions of the ``easiest'' ferroelectric
transformation path and to clarify the role of Bi ions on the
ferroelectric distortion. The potential energy surface associated
with the displacement of Bi ion along all the three directions
(see Fig.\,\ref{fig:dist}a) are having a double well shape. From
Fig.\,\ref{fig:dist} it is clear that the lowest energy off-center
displacements are along the [111] directions. This confirms that
the polarizability of Bi plays a special role in the ferroelectric
properties. The electronic structure studies on
Pb(Zr$_{0.5}$Ti$_{0.5}$)O$_3$ show~\cite{wu03} polarization
rotation from the [100] to the [111] direction owing to the
smaller energy difference between the tetragonal and the
rhombohedral structures. An experimental study~\cite{guo00} on PZT
showed that the piezoelectric elongation in tetragonal PZT is
along the direction associated with a monoclinic distortion,
suggesting the mechanism of polarization rotation via the
monoclinic phase. Our total energy studies on BiFeO$_3$ show that
the energy difference between the Bi ion displacement along [111]
and [110] directions are relatively small.  Consistent with our
observation, experimental studies~\cite{wang03} indicate that
heteroepitaxially strained thin films of BiFeO$_3$ have
tetragonal-like crystal structure with the $c$ axis normal to the
substrate surface, with a small monoclinic distortion of about
0.5$^{o}$.

\par
Figure\,\ref{fig:dist} shows that the energy gain associated with
the ferroelectric distortion along [111] direction is 0.427~eV per
f.u. This energy gain is found to be much higher than 0.056\,eV
for PZT, the well known piezoelectric material~\cite{iniguez03}.
It is tempting to try to connect these numbers with the
corresponding ferroelectric Curie temperatures. We know that, in
principle, the thermal energy required for a system to remain in
the high-symmetric paraelectric $R$\={3}c phase will be of the
order of energy barrier between these two systems. This suggests
that the the Curie temperature of BiFeO$_3$ must be much higher
than that of PZT. Consistent with this observation the
ferroelectric Curie temperature of BiFeO$_3$ is ($\sim$ 1000~K)
larger than that of  PZT ($\sim$ 570~K). It should be noted that
for systems with a competition between different structural
instabilities~\cite{halilov02}, such an argument can grossly
overestimate the transition temperature. In BiFeO$_3$, for
example, it is known that there is a competition between the
ferroelectric rhombohedral phase (arising from the off-center
displacements of ions due to Bi lone pairs) and an
antiferro-distortive phase (associated with the lattice mismatch
involving rotation of the oxygen octahedra).

\subsection{Chemical Bonding}
\begin{figure*}
\includegraphics[scale=0.65,angle=0]{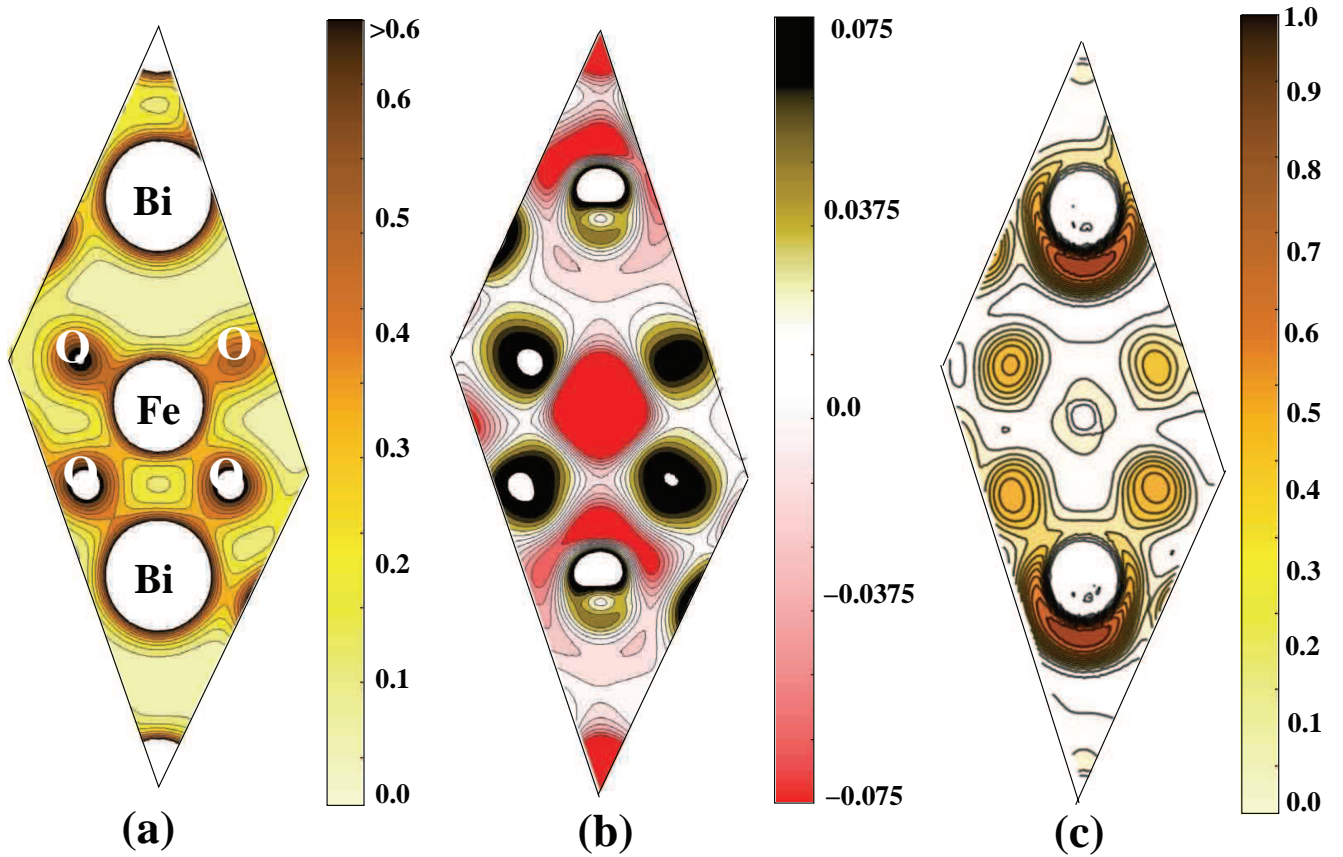}
\caption{(Color online) Calculated (a) valence-electron charge
density, (b)charge-transfer  and (c) ELF plot for ferroelectric
BiFeO$_3$.} \label{fig:charge}
\end{figure*}

 The ferroelectric properties of materials such as
Pb(Zr$_{1-x}$Ti$_{x}$)O$_{3}$ are known to be related to the
partial covalency of some bonds~\cite{cohen92,posternak94}. In
this connection detailed analysis of chemical bonding in BiFeO$_3$
is particularly interesting. Analysis of bonding interaction from
charge density distribution alone may mislead the identification
of the nature of chemical bonding correctly~\cite{raviprl}. Hence
we have used different tools to explore the bonding behavior in
BiFeO$_3$.
 From the analyses of the charge
density and densities of states for PbTiO$_3$, and BaTiO$_3$ it is
concluded~\cite{cohen90,cohen92} that the ferroelectric
instability is due to hybridization between the O 2$p$ states and
the Ti 3$d$ states. So, it is interesting to analyze the chemical
bonding between Fe and O in BiFeO$_3$ in order to have a better
understanding on the origin of ferroelectricity. The electron
densities calculated by VASP were shown for a plane containing Bi,
Fe and O atoms in Fig.\,\ref{fig:charge}a. From this figure it is
clear that the bonding interaction between Fe and O is not purely
ionic. The directional nature of charge density distribution
between Fe and O indicates the presence of finite covalent
bonding.

\par
Experimental results~\cite{nemoshkalenko85} and detailed
theoretical investigations~\cite{ghosez95} on BaTiO$_3$ showed the
significant hybridization interaction between Ba 5$p$ and O 2$p$
orbitals which contributes to the Born effective charge and
ferroelectric properties. In this context, the analysis of the
nature of bonding interaction between Bi and O is particularly
interesting.  The charge density distribution in
Fig.\ref{fig:charge}a shows presence of finite charges between Bi
and O, which is generally an indication for the covalent
interaction. Charge transfer distribution (see
Fig.\ref{fig:charge}b) shows that electrons transfer from both Bi
and Fe atoms to the oxygen sites consistent with the ionic
picture. If one has pure ionic bonding interaction between Bi and
O as well as between Fe and O, one would expect an isotropic
charge transfer distribution. The anisotropic distribution of
charge transfer confirms the presence of finite covalent bonding
between Bi and O as well as between Fe and O.

\par
ELF is an informative tool to distinguish different bonding
interactions in solids~\cite{savin97}, here it is shown for
ferroelectric BiFeO$_3$ in Fig.~\ref{fig:charge}c. The negligibly
small value of ELF between atoms indicate the presence of dominant
ionic bonding in this material. Also the ELF distribution given in
Fig.~\ref{fig:charge}c shows maximum value at the O sites and
minimum value at the Fe and Bi sites reiterate charge transfer
interaction from Bi/Fe to O sites. Moreover polarization of ELF at
the O sites towards the other O sites and finite ELF between Bi
and O indicate the hybridization interaction. The conclusion from
the charge density, charge transfer and ELF analyses is that the
bonding interaction between Bi and O as well as Fe and O are
dominant ionic bonding with finite covalent character. This
conclusion is consistent with the conclusions arrived from the
Born effective charge analysis discussed below. Our results
clearly establish the mixed ionic-covalent character of the
bonding in BiFeO$_3$: it is obvious that the covalent character is
not restricted to the Fe$-$O bond. It may be noted that while Bi
$s-$O $p$ bonding is not very strong, it is at least as important
as the Fe $d-$O $p$ interaction.

 \par
If the bonding between the constituent atoms in  BiFeO$_3$ is
purely ionic it would remain centrosymmetric (and therefore not
ferroelectric), because the short-range repulsion between adjacent
closed-shell ions is minimized for symmetric structures. The
existence or absence of ferroelectricity is determined by a
balance between these short-range repulsions, which favor the
non-ferroelectric symmetric structure, and additional bonding
considerations which act to stabilize the distortions necessary
for the ferroelectric phase. Until now two different chemical
mechanisms have been identified for stabilizing the distorted
structure in common ferroelectrics. In the first mechanism, the
ferroelectric distortion was ascribed to the hybridization
interaction between the transition metal ion and
oxygen~\cite{cohen92}. The transition metal ions (such as
Ti$^{4+}$, Zr$^{4+}$, Nb$^{5+}$ and Ta$^{5+}$) participating in
hybridization interaction with O are formally in a $d^{0}$ state,
hence the lowest unoccupied energy levels are $d$-states which
tend to hybridize with O $2p$ states~\cite{hill00}. However, this
mechanism is not very significant for the case of multiferroics
because the transition metal ions are not in a $d^{0}$ state.

\par
 The second mechanism involves cations with ``lone pair"
electrons which have a formal $ns^{2}$ valence electron
configuration. In the same vein as the $d^{0}$ ions discussed
above, these $p^{0}$ ions (Tl$^{+}$, Sn$^{2+}$, Pb$^{2+}$,
Sb$^{3+}$, and Bi$^{3+}$) contain some $p$ charge density which
contribute to the displacive distortions. Indeed the tendency of
$ns^{2}$ ions to lose inversion symmetry is well established, with
the conventional explanation invoking a mixing between the
localized $ns^{2}$ states and a low-lying $ns^{1}np^{1}$ excited
state, a mixing that can only occur if the ionic site does not
have inversion symmetry~\cite{atanasov01}. If the lowering in
energy associated with the hybridization interaction is larger
than the inter-ionic repulsion opposing the ion shift, then a
ferroelectric distortion results. This "stereochemical activity of
the lone pair'' is the driving force for off-center distortion in
magneto-electric materials.

\subsection{COHP}
The crystal orbital Hamiltonian population (COHP) is calculated
using the tight binding linear muffin-tin orbital program
(TBLMTO)~\cite{andersen84}. The COHP is the density of states
weighted by the corresponding Hamiltonian matrix element and is
indicative of the strength and nature of a bonding (negative COHP)
or antibonding (positive COHP)
interaction~\cite{dronskowski93,cohp}.

\begin{figure}
\includegraphics[scale=0.5]{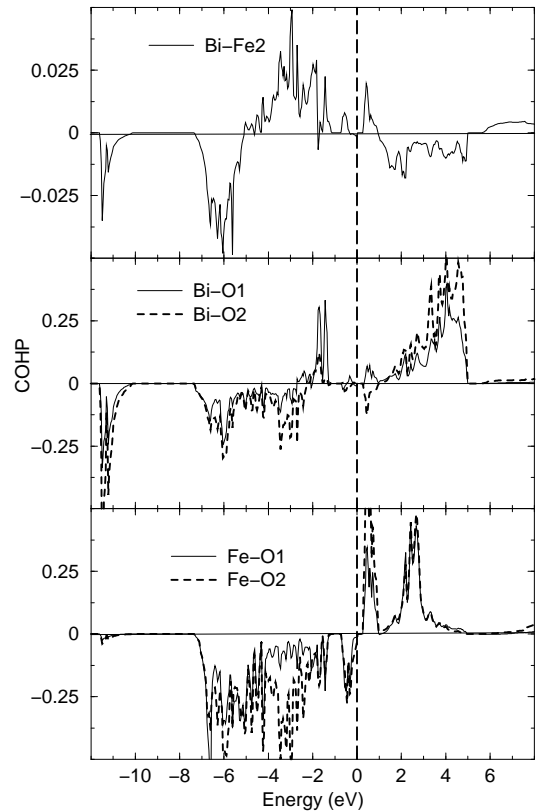}
\caption{ COHP for BiFeO$_3$ in the ferroelectric $R3c$ structure
describing Bi$-$O, Fe$-$O and Bi$-$Fe interactions. O1 and O2 are
the planar and apical oxygen atoms, respectively.}
\label{fig:cohp}
\end{figure}

\par
The calculated COHP for the Bi$-$O, Bi$-$Fe, and Fe$-$O
interaction for ferroelectric BiFeO$_3$ are shown in
Fig.\,\ref{fig:cohp}. The COHP curves reveal that all occupied
states for the Fe$-$O interaction have bonding states, and the
Fermi level is perfectly adjusted to fill up these bonding states.
So the strongest bonding interaction is between Fe and O atoms.
The FeO$_6$ octahedra are highly distorted and hence the Fe$-$O2
bonding energy ($-$1.52~eV) is stronger than the Fe$-$O1
interaction ($-$1.02~eV) where O1 and O2 are planar and apical
oxygen atoms, respectively. Considerable amount of bonding states
within the valence band (VB) in COHP shows significant bonding
interaction between Bi and O atoms.
It should be noted that a large bonding-state peak appearing
around $-$10~eV (between the Bi-O interaction) is due to the
hybridization interaction between the Bi 6$s$/6$p$ and O 2$s$
states. As the COHP is estimated from TBLMTO method, we found a
shift in the DOS value by around 1~eV towards higher energy with
respect to that from VASP. From the COHP analysis it is clear that
the hybridization interaction between Bi/Fe and O is significant,
and it is an important driving force for the ferroelectric
distortion. The COHP for Bi$-$Fe interaction shown in
Fig.\,\ref{fig:cohp} indicates the presence of both bonding and
antibonding states within the VB. However, the bonding interaction
between Bi and Fe in BiFeO$_3$ is weaker than that between all the
other atoms.

\subsection{Band structure results}
\begin{figure*}
\begin{minipage}{\textwidth}
\hspace{-0.60in}
\includegraphics[scale=0.55]{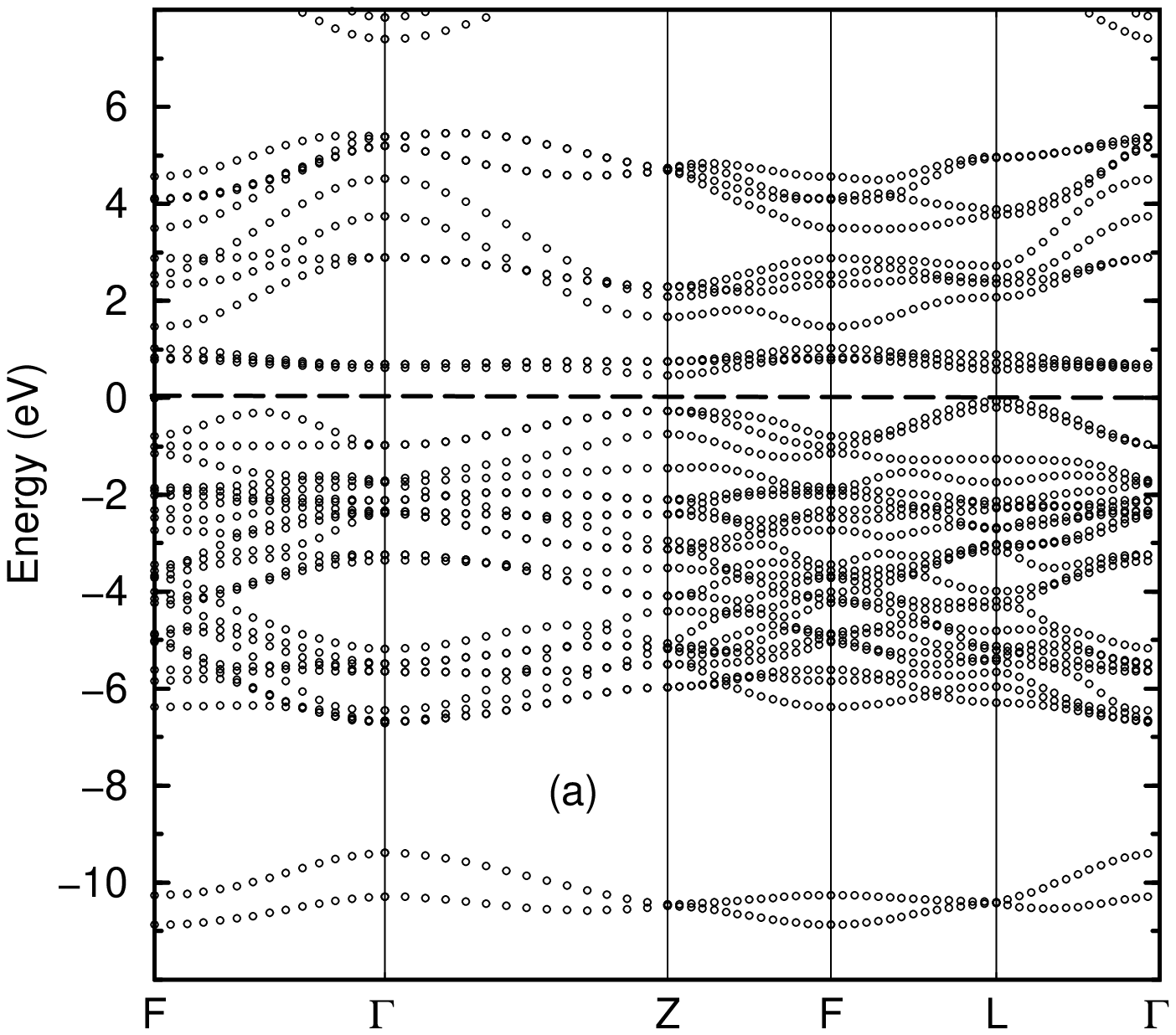}
\includegraphics[scale=0.55]{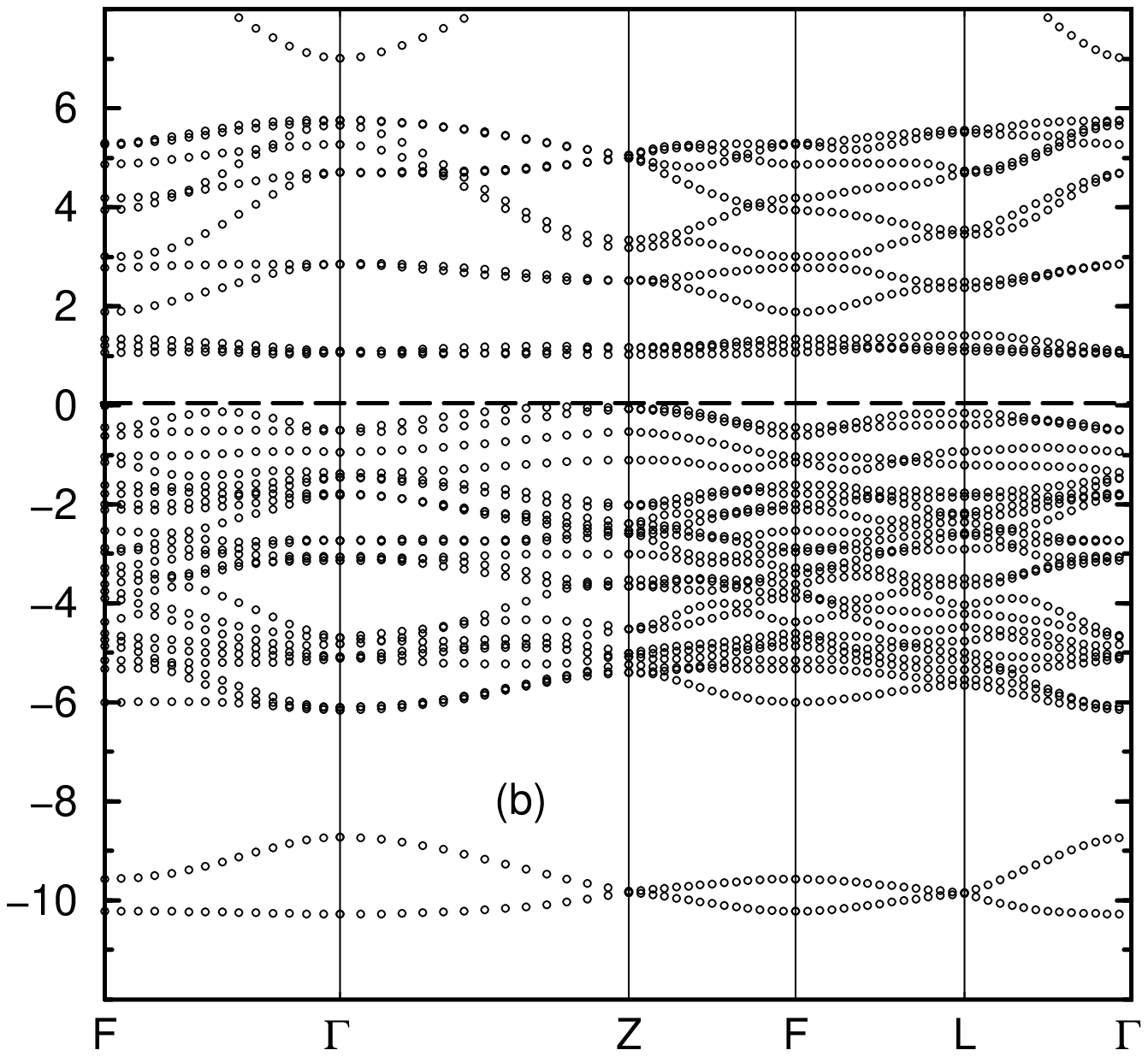}
\caption{\label{fig:bnds}Electronic band dispersion for
antiferromagnetic BiFeO$_3$ in (a) paraelectric $R$\={3}$c$ and
(b) ferroelectric $R3c$ structures. The Fermi level is set to
zero.}
\end{minipage}
\end{figure*}
\par
\begin{figure}
\includegraphics[scale=0.5]{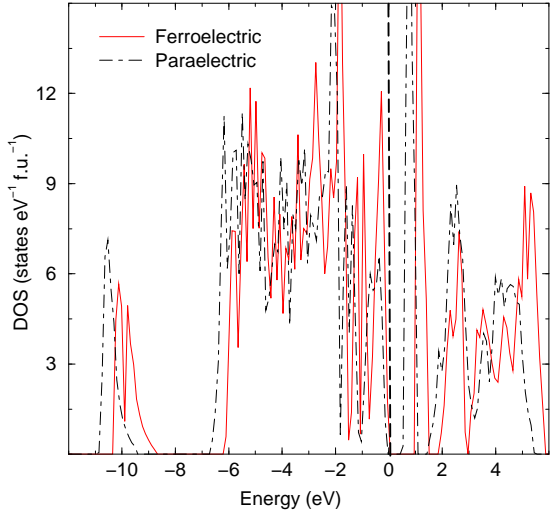}
\caption{(Color online) Calculated total density of states for
BiFeO$_3$ in the paraelectric $R$\={3}$c$ and ferroelectric $R3c$
phases.} \label{fig:fpdos}
\end{figure}

In order to gain more insight into the role of ferroelectric
distortion on electronic structure and bonding behavior in
BiFeO$_3$, the calculated band structure for A-AF phase of
BiFeO$_3$ in the paraelectric $R$\={3}$c$ structure and the
ferroelectric $R3c$ structure are given in Figs.\ref{fig:bnds}a
and b, respectively.
As in R\={3}c phase (Fig.\,\ref{fig:bnds}a), the highest-energy
valence states have noticeable O 2$p$ character and the lowest
conduction band states are Bi 6$p$-like, at variance with the Ti
3$d$ bands in PbTiO$_3$. This together with the relatively narrow
band gap of the paraelectric R\={3}c phase, results in
ferroelectric instabilities mainly characterized by a
hybridization between Bi 6$p$ and O 2$p$ orbitals, not the one
involving $d$ bands of the transition metal cation that is typical
of most other ferroelectric
perovskites~\cite{posternak94,kingsmith94}. It is interesting to
note that the highest energy valence band state of $R3c$ phase has
dominant Fe 3$d$ states and these states are shifted to around
0.35~eV to the lower energy in the paraelectric phase. In the
$R3c$ phase (see Fig.\,\ref{fig:bnds}b) the gap opens considerably
and the top valence bands increase their Bi 6$p$ character (See
also Fig.\,\ref{fig:fpdos}). The strong Bi 6$p-$O 2$p$
hybridization is also reflected in the fact that the effective
charge on Bi in the ferroelectric R3c phase has larger value than
that in the paraelectric R\={3}c phase (see
Section\,\ref{sec:charge}).
\par
Two well dispersed bands around $-$10~eV is mainly originating
from Bi-6$s$ electrons. Compared with the paraelectric phase,
these two bands are slightly broader in ferroelectric phase.  The
O 2$s$ bands are completely filled and they are located around
$-$18~eV below the Fermi level and are not shown in
Figs.~\ref{fig:bnds}a and b. The parabolic nature of bands around
$-$6~eV are from Fe 4$s$ electrons with significant contribution
from both O 2$p$ and Fe 3$d$ electrons. A series of bands between
$-$6~eV and the Fermi level are mainly originating from hybridized
bands of both Fe 3$d$ and O 2$p$ electrons.
\par
Owing to the off-center displacement of the ions in the
ferroelectric phase of BiFeO$_3$, the degeneracy in bands along
the $F-\Gamma$ direction in Fig.\,\ref{fig:bnds}b are lifted
compared to that of paraelectric phase. When the system is allowed
to relax to its magnetic ground state a gap opens and an insulator
is formed, as required for the ferroelectric state. The
ferroelectric distortion slightly increases the band gap values,
from 0.46~eV in the paraelectric to 0.97~eV in the ferroelectric
phases.
The density functional calculations are known to underestimate the
band gap and hence the reported values will be smaller than the
real band gap value for this material. The Bi ions are almost in
the $3+$ ionic state and hence the contribution from Bi 6$p$
electrons are relatively small in the VB and they are mainly in
the CB around 2 to 6\,eV.

\begin{figure}
\includegraphics[scale=0.5]{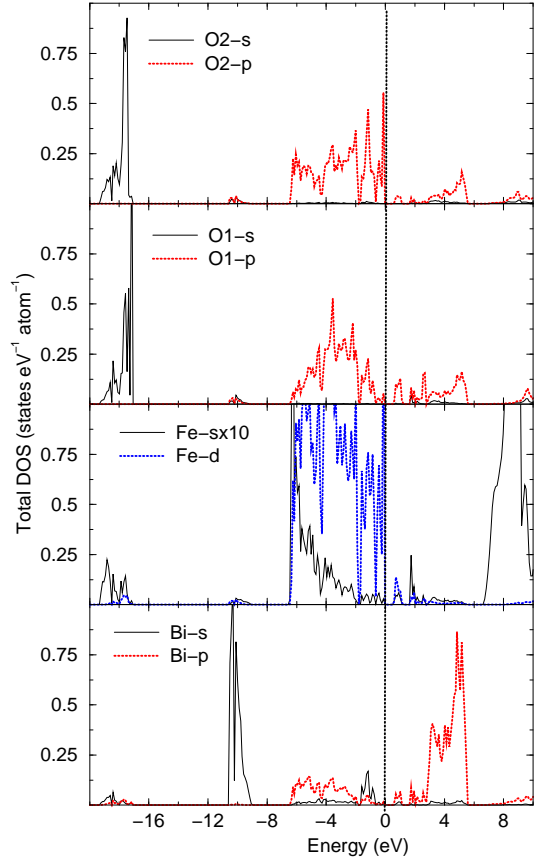}
\caption{(Color online) Partial density of states for
ferroelectric BiFeO$_3$ in the G-AF state.} \label{fig:pdos}
\end{figure}

\par
In order to understand the sensitivity of electronic structure by
ferroelectric displacement, the calculated total density of states
for the paraelectric and the ferroelectric phases of BiFeO$_3$ are
superimposed in Fig.~\ref{fig:fpdos}.  From this figure it is
clear that overall the energy of electrons in the ferroelectric
phase are shifted around 0.5\,eV from the paraelectric phase due
to the off-center displacement of ions. This also highlights the
different bonding interaction between planar oxygen and Fe due to
the buckling of the FeO$_6$ octahedra to accommodate the Bi 6$s^2$
lone pair electrons occurring after the phase transition.
In fact, the changes in the ferroelectric phase found at the top
of the VB involve the $d_{z^{2}}$ states of Fe and the $p_z$
states of the apical oxygen, whose stronger hybridization gives
more stability to the structure in the ferroelectric phase.
\par
To simplify the interpretation of the electronic structure, we
have also calculated the partial (atom and orbital decomposed)
densities of states for the ferroelectric BiFeO$_3$ (see
Fig.~\ref{fig:pdos}). The top of  VB is set to zero. Around
$-$18\,eV the lowest energy VB states form a filled, rather narrow
band from O $2p$ electrons. Above these states (around $-$10~eV)
lone pair Bi 6$s$ states with small contribution from O 2$s$ and
2$p$ electrons are present. The Fe $4s$ states are located around
$-$6 eV in the VB and they are completely mixed with the Fe $3d$
states in the entire VB. The energetically degenerate nature of Fe
$3d$ and O $2p$ in the whole VB indicates the presence of covalent
bonding between these two atoms. Bi 6$s$ and 6$p$ states are well
separated and the latter states are mostly in the CB implying
highly ionic nature of Bi ions.


\subsection{Magnetic properties}
Crystalline BiFeO$_3$ is a strong antiferromagnet (G-type spin
structure, $T_{\rm N}$ = 654.9~K)~\cite{blaauw73}. The cooperative
magnetism in BiFeO$_3$ is originating from the half-filled and
localized ($t_{2g}^{3}e_{g}^{2}$) Fe$^{3+}$ ions. In an ideal
cubic perovskite-like lattice, the Fe$^{3+}$ ions in the high spin
state prefer to form the G-AF ordering because, in this case the
Pauli exclusion principle allows the transfer of electron to the
neighboring ion in an antiparallel direction only.  The neutron
diffraction studies carried out at room temperature revealed that
BiFeO$_3$, in fact, exhibits a compensated antiferromagnetic
ordering ($T_{\rm N}$ = 670~K), with a cycloidal spin arrangement,
incommensurate with its lattice~\cite{sosnowska821}. The magnetic
structure of a separate unit cell corresponds to the G-type
antiferromagnetic ordering (each magnetic Fe$^{3+}$ ion is
surrounded by six Fe$^{3+}$ ions with  spins directed opposite to
that of the central ion)~\cite{kiselev63}. Indirect evidence of
the existence of the spatially spin-modulated structure (SSMS) in
BiFeO$_3$ was obtained~\cite{popov93} from the measurements of the
magnetoelectric effect. The existence of SSMS with periodicity of
620~{\AA} was recently confirmed by $^{57}$Fe NMR measurements
~\cite{zalesskii02}.
\par
Dzyaloshinskii~\cite{dzyaloshinskii57} developed a thermodynamic
theory based on spin-lattice interaction and magnetic dipole
interaction to explain the origin of weak ferromagnetism in
antiferromagnetic compounds. The basic assumption is that
antiferromagnetically ordered spins could rotate about a crystal
axis toward one another, resulting in a net spontaneous moment
over the unit cell. The calculated magnetic moments and the total
energy with respect to the ground state magnetic configuration for
BiFeO$_3$ are given in Table\,\ref{table:de}. It is interesting to
note that the energy difference between the ferromagnetic state
and different antiferromagnetic state is larger compared with that
for LaMnO$_3$ (in LaMnO$_3$ the energy difference between the
ferromagnetic and the A-AF state is only 25
meV/f.u.)~\cite{ravilamn}. This observation is consistent with the
experimental measurements in the sense that amorphous phase of
BiFeO$_3$ is found to be in a speromagnetic state (short range
antiferromagnetic behavior) rather than in a ferromagnetic
state~\cite{nakamura93}.

\par
Like LaFeO$_3$, BiFeO$_3$ also has Fe ions in the $3+$ states.
Similar to LaFeO$_3$ one could expect G-AF ordering in BiFeO$_3$.
But, owing to the presence of lone pair electrons at the Bi sites,
one could expect anisotropy in the exchange interaction also.
Consistent with this expectation we found that the A-AF state is
found to be 171~meV/f.u. lower in energy than the G-AF state in
BiFeO$_3$. This suggests that the Bi lone pair electrons play an
important role in not only  the ferroelectric behaviors but also
in the magnetic properties of BiFeO$_3$. The observation of A-AF
ordering is particularly interesting. In LaMnO$_3$ each MnO$_{6}$
octahedron is elongated along one axis as a result of the
Jahn$-$Teller distortion around the $d^{4}$ Mn ions. This
Jahn-Teller distortion creates orbital ordering that leads to four
singly occupied $d_{z}^{2}$ orbitals pointing towards empty
$d_{x^{2}-y^{2}}$ orbitals on adjacent Mn ions (a ferromagnetic
interaction)  and two empty $d_{x^{2}-y^{2}}$ orbitals pointing
towards other empty $d_{x^{2}-y^{2}}$ orbitals (an
antiferromagnetic interaction) resulting in the A-AF
ordering~\cite{ravilamn}. If no other mechanism was involved in
the stabilization of magnetic ordering in BiFeO$_3$ one could
expect G-AF ordering as that in LaFeO$_{3}$. But, owing to the Bi
lone pair electrons, we found A-AF phase to be the ground state
for BiFeO$_3$ among the considered collinear magnetic
configurations. It should be noted that if the Coulomb correlation
effects are important in this material, one could expect that the
energetics of different magnetic configurations will be different
from the values reported here. But, as most of the features of
magnetoelectric behavior of BiFeO3 are well described by GGA
itself, we have not considered Coulomb correlation effects in the
present study.

\begin{table}
\caption { Total energy (relative to the lowest energy state in
meV/f.u.), Fe-magnetic moment (in $\mu_{B}/atom$), and total
magnetic moment (including those at oxygen and interstitial sites;
$\mu_{B}$/f.u.) for BiFeO$_3$ in A-, C-, G-AF, and F states in the
ferroelectric $R3c$ structure. }
\begin{ruledtabular}
\begin{tabular}{l c c c l}
                &      A-AF  &      C-AF  &       G-AF    &     F  \\
\hline
Energy          &      0     &      211   &       171     &   390 \\
Fe-Magnetic moment &    3.724 & 3.824 &  3.789   &  3.936\\
Total magnetic moment & 3.885 & 3.974 &  4.076   &  4.893 \\
\end{tabular}
\end{ruledtabular}
\label{table:de}
\end{table}

The calculated magnetic moment at the Fe site varies between
3.724$-$3.936~$\mu_{B}$ per Fe atom depending upon the magnetic
configuration considered in our calculations (see
Table\,\ref{table:de}). The calculated magnetic moment at the Fe
site is found to be in good agreement with the value of
3.75(2)$\mu_{B}$ measured from low temperature neutron diffraction
measurements~\cite{sosnowska02}. The calculated magnetic moments
at the iron sites are not integer values, since the Fe electrons
have a hybridization interaction with the neighboring O ions.
Because of this hybridization interaction we found that around
0.18~$\mu_ {B}$ magnetic moment is induced at each O site in the
ferromagnetic state which are polarized along the same direction
as that in the Fe sites. As a result the net total moment at the
site is considerably larger (see the total moment in
Table\,\ref{table:de}).

\begin{figure}
\includegraphics[scale=0.5]{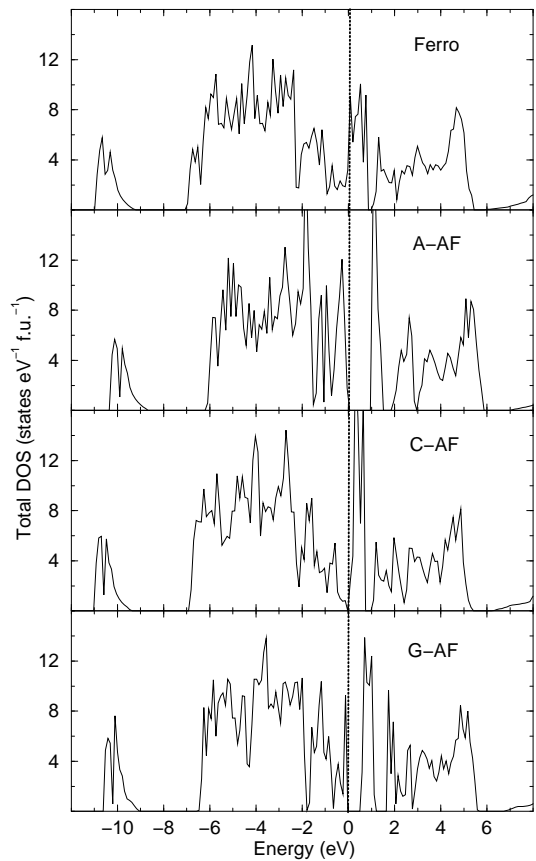}
\caption{Calculated total density of states for BiFeO$_3$ in different magnetic
configurations. The Fermi level is set to zero.}
\label{fig:tdos}
\end{figure}
\par
The calculated total DOS for BiFeO$_3$ in the nonmagnetic and
ferromagnetic states show metallic behavior with a large DOS at
the Fermi level. The large number of electrons near the Fermi
level is not a favorable condition for stability. However, when
one creates AF ordering, the exchange interaction produces an
exchange potential that effectively shifts the energy of the Fe
3$d$ band to lower energy giving insulating behavior as shown in
Fig.~\ref{fig:tdos}. We found that even if we introduce spin
polarization into the calculation, the electronic ground state
depends upon the nature of magnetic ordering. The Fermi level lies
in a pseudogap feature in the C-AF state and clear insulating
behavior is observed for the A-AF and G-AF ordered states with
band gaps of 0.97~eV and 0.53~eV, respectively. If the correlation
effects are not significant in this material, the present result
suggest that magnetic ordering plays an important role in
stabilization of ferroelectricity. However, it should be noted
that the magnetic ordering temperature is much smaller (643 K)
than the ferroelectric ordering temperature (1100 K). Hence, it
appears that mechanisms other than magnetism may establish
insulating behavior and ferroelectricity at high temperatures.


\subsection{Role of Bi lone pair electrons for ferroelectricity}

\begin{figure}
\includegraphics[scale=0.5]{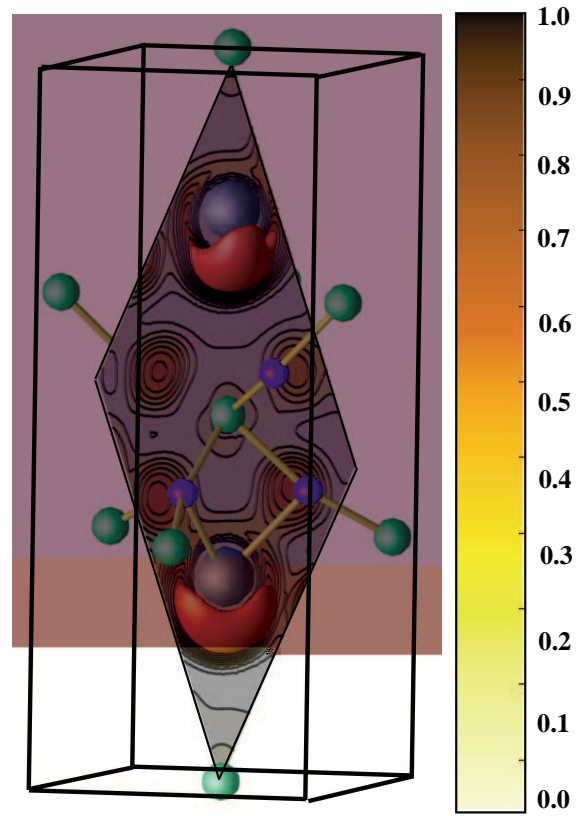}
\caption{(Color online) Isosurface (at a value of 0.75) of the
valence electron localization function of BiFeO$_3$ in the
ferroelectric $R3c$ structure obtained from TBLMTO calculation.
The atom labels in the shown plane are same as that in
Fig.~\ref{fig:charge}a} \label{fig:elf}
\end{figure}

The ELF provides a measure for the local influence of the Pauli
repulsion on the behavior of electrons and permits the mapping in
real space of core, bonding, lone pair, and nonbonding regions in
a crystal~\cite{silvi94}. We have used the TBLMTO program to
calculate the ELF. In order to visualize the lone pairs around Bi,
it was necessary to select an isosurface with an ELF value of 0.75
as shown in Fig.~\ref{fig:elf}. This figure reveals the lobe-like
lone pairs possessed by the Bi atoms. It also indicates that the
Bi 6$s$ orbitals are in direct interaction with oxygen 2$p$ and
are thus not chemically inert. An interesting feature of ELF for
BiFeO$_3$ in Fig.~\ref{fig:charge}c is the presence of a
nonuniform distribution of ELF at the Bi sites. The reason for
this is that due to the displacement of the Bi atoms, the Bi
6$p_{z}$ orbital can interact more efficiently with the O 2$p_{x}$
and O 2$p_{y}$ orbitals. In the R\={3}c structure, these orbitals
interact only with the Bi 6$s$ orbitals because the symmetry does
not allow for an interaction with the Bi 6$p_{z}$ orbitals. In the
$R3c$ structure, half of the Bi atoms have moved in the $+z$
direction and another half of them in the $-z$ direction, breaking
the symmetry and allowing hybridization between the indicated
linear combination of O orbitals and the Bi $p_{z}$ orbital.  As a
result, some of the O $p_{x}$/O $p_{y}$ derived states (in fact
those that had the strongest interaction with the Bi 6$s$ state)
are lowered in energy and move away from the Fermi surface,
effectively counterbalancing the destabilization caused by the Bi
lone pair. The combination of large values of ELF and the
covalency leads to a stronger lone pair, which is more
stereochemically active.  The manner in which the lone pairs
dispose themselves suggests a greater tendency to the
ferroelectric distortion.

\par
The lone pairs are expected to have primarily 6$s$ character. If
they are to form lobes, there must be some hybridization
interaction with the neighbors. The question is whether the $p$
character arises from mixing $s$ and $p$ states on the Bi atom or
 is it an effect of covalency, arising from Bi $s$ and O
$s/p$ mixing. The partial DOS analysis (see Fig.\,\ref{fig:pdos})
shows that Bi 6$s$ electrons are around $-$10~eV below the Fermi
level and are well localized. Also the Bi-$s$ and $p$ states are
well separated indicating that the mixing between $s$ and $p$ on
the Bi site is minimal. Watson and Parker~\cite{watson99} have
recently suggested that it is due to $p$ character from the anions
that lone pairs are allowed to form lobe-shaped structures. The
anion is clearly implicated since in a homologous series of
compounds, changing the anion often greatly affects the
stereochemistry of the lone pair~\cite{seshadri01}. If the lone
pair 6$s$ electrons overlap with the O $s/p$ states it can attain
the typical lobe shape, conversely, in a situation where there is
no such overlap, or when the overlap is weak, the lone pair could
retain its pure 6$s$ character and remain spherical.

\par
 In BiFeO$_3$ the charge density and ELF analysis (see Fig.\,\ref{fig:charge})
show noticeable covalent bonding between Bi and O.  In order to
have covalent bonding between Bi $s$ and O $p$ states they should
be energetically degenerate. On the other hand the O $p$ states
are mainly present at the top of the VB. Although both Bi $s$ and
O $s$ states are well localized and energetically well separated,
they are spatially well dispersed. Owing to the spherically
symmetric nature of the $s$ orbitals, the probability of the
hybridization interaction between $s$ orbitals is much larger than
that between any other orbitals.  Hence the lobe shape of the lone
pair is at least partly associated with the hybridization
interaction between Bi $s$ and O $s$ states. Even though O $p$
states are mainly present at the top of the valence band, finite O
$p$ electrons are present around $-$10\,eV where the Bi 6$s$
electrons are also present. So, both O $s$ and $p$ electrons
participate in the hybridization interaction with the Bi 6$s$
electrons. Consistent with this view point our COHP analysis (see
Fig.\,\ref{fig:cohp}) indicates that Bi$-$O bonding states are
present around $-$10\,eV.

\par
It was argued by Orgel~\cite{orgel59}, and has been assumed in
much of the literature, that the interaction between occupied
cation $s$  and unoccupied cation $p$  bands is the primary cause
for the lobe formation. The partial DOS analysis clearly shows
(see Fig.\,\ref{fig:pdos}) that Bi $s$ and $p$ states are well
separated. Hence our observation does not support this classical
view of hybridization of the Bi 6$s$ and 6$p$ orbitals. Rather, it
supports  the observation of hybridization interaction between the
anion $s$ and cation $p$ states as seen in PbO and related
compounds~\cite{watson99,seshadri01}. The formation of the
distorted electron densities and crystal structures will thus
depend on the energies of the valence orbitals of oxygen ions
involved in bonding to the Bi.

\par
Although the $s$ states of both Bi and O are well below the VB
their importance for ferroelectric distortion can be understood as
follows. The $s$ states are spherically symmetric and diffuse
compared to the $p$ or $d$ states. Hence, the overlap interaction
between $s$ states will be stronger than that between other
states. Owing to the highly ionic nature of Bi, the Bi 6$p$ states
are mostly in the unoccupied state (see Fig.\,\ref{fig:pdos}) but
with some noticeable Bi 6$p$ states also at the VB. It has been
shown that the stereochemical activity of the cation lone pair is
determined by the interaction of cation $s$ and $p$ states with
anion $p$ states~\cite{waghmare03}. Hence the activity of the lone
pair will be stronger if the covalent interaction between Bi and O
is larger, where it has the stereochemical effect of stabilizing
the distorted low symmetric structures. Stability of a distorted
structure depends on whether the energy gained from hybridization
interaction (bonding) between occupied and unoccupied states
compensates the energy required to promote an electron to
unoccupied state. Since the same oxygen $p$ states are involved in
both interactions and hence in instabilities, a competition
results.

\par
BiFeO$_3$ belongs to the ferroelectrics of displacive type. Now we
try to understand the role of lone pair on the ferroelectric
behavior. The lone pair occupies as much volume as an oxide or
fluoride anion in crystals~\cite{hyde89}. The 6$s^{2}$ electrons
of Bi$^{3+}$ hybridize with 2$s$ and 2$p$ of oxygen to form a
space-filling localized lobe, which in turn pushes away its
neighboring atoms causing a structural distortion. The presence of
lone pair makes the hybridization interaction between Bi and O
different in different directions. Moreover, the effective ionic
radius of Bi ions decreases on one side and increases on the other
side. The anisotropic nature of electrons at the Bi sites
(originating from the presence of lone pair) stabilizes off-center
displaced structure.  This results in the movement of the Bi ion
in the [111] direction which causes a cooperative displacement of
the Fe$^{3+}$ ions due to the repulsive interaction and finally
gives rise to a ferroelectric polarization.

\subsection{Born effective charge analysis}
\label{sec:charge}

 The Born effective charge (BEC) $Z^{*}$ is a fundamental
quantity for the study of lattice dynamics, because it governs the
amplitude of the long-range Coulomb interaction between nuclei,
and the splitting between longitudinal optic (LO) and transverse
optic (TO) phonon modes.  In binary crystals~\cite{phillips70},
infra-red measurements of the splitting between LO and TO modes
allow an accurate estimation of $\|Z^{*}\|^{2}/\epsilon_{\infty}$
and offer therefore an unambiguous way to extract the amplitude of
$Z^{*}$ from the experiment. But, in complex compounds like
perovskite oxides, LO and TO mode eigenvectors are not necessarily
equivalent and the determination of $Z^{*}$  from the experimental
data is consequently not straightforward~\cite{axe67}. Hence, the
development of theoretical methods giving direct access to $Z^{*}$
acquires specific interest.

\par
The dynamical charge  measures the macroscopic current flowing
across the sample while the ions are adiabatically displaced. Such
currents are responsible for building up the spontaneous
polarization, when ions are continuously displaced from the
centrosymmetric structure to the ferroelectric
structure~\cite{resta92,kingsmith93}. These displacements are
directly observable when measuring polarization via hysteresis
cycles. The modern theory of macroscopic
polarization~\cite{resta94} allows the calculation of such
currents as Berry phases. The dynamical Born effective-charge
tensor $Z^{*}$ of a given ion measures (by definition) total
macroscopic current adiabatically transported by the ion moving
with unit velocity, while all the other ions and the macroscopic
field are kept fixed~\cite{pick70}.
\par
According to the modern theory of
polarization~\cite{kingsmith93,vanderbilt94,resta94}, the change
in macroscopic polarization between two different insulating
states can be regarded as a measure of the phase difference
between their initial and final many-body wavefunctions. The total
polarization {\bf P} for a given crystalline geometry can be
calculated as the sum of ionic and electronic contributions.
\par
The ionic contribution is obtained by summing the product of the
position of each ion in the unit cell with the nominal charge of
its rigid ion core. The electronic contribution to {\bf P} is
determined by evaluating the phase of the product of overlaps
between cell-periodic Bloch functions along a densely sampled
string of neighboring points in {\bf k} space.
\par
It has been shown~\cite{resta93,zhong94} that the effective-charge
tensors are almost constant along the path connecting the two
structures, which implies that the spontaneous polarization is
simply expressed in terms of the internal-strain displacement
$u_{\tau}$ as

\begin{eqnarray}
\label{eqn:pol} {\bf P}_{\tau} = \frac{\big | e \big
|}{\Omega}\Sigma_{\tau}Z^{*}_{\tau}u_{\tau},
\end{eqnarray}

where $\Omega$ is the cell volume, and $Z^{*}_{\tau}$ are the
diagonal components of the effective-charge tensors for atom
$\tau$. BEC can be defined as a linear term in the polarization
change due to a unit displacement of the $\tau$th ion, keeping all
the other ions fixed under the condition of zero macroscopic
field~\cite{pick70}. i.e.,
\begin{eqnarray}
\label{eqn:z} Z^{*}_{\tau\alpha\beta} = \Omega\frac{\Delta {\bf
P}_{\beta}}{\Delta u_{\tau\alpha}}\big| _{\epsilon = 0}
\end{eqnarray}
for the change in $\bf{P}$ on moving ion $\tau$; note that the
effective charge is a tensor. Thus the effective charge is
calculated by finding the induced polarization of the nuclei and
the electrons when the nuclei are displaced equally in each unit
cell, so that the translational periodicity is the same as the
undistorted crystal. The effective charge thus found is the
transverse Born effective charge~\cite{lax55,born54} since the
macroscopic electric field is required to vanish. According to the
modern theory of polarization~\cite{resta94,kingsmith93}, the
total difference in polarization between a reference state and a
state where the atoms belonging to the sublattice $\tau$ have been
displaced by a small but finite distance $\Delta u_{\tau\alpha}$
is defined as,
\begin{eqnarray}
\Delta{\bf P} = \Delta{\bf P}_{el} + \Delta{\bf P}_{ion}
\end{eqnarray}

where $\Delta{\bf P}_{el}$ is the electronic contribution obtained
from the Berry-phase polarization approach and $\Delta{\bf
P}_{ion}$ is the ionic contribution.  $\Delta{\bf P}_{ion}$ is
defined by

\begin{eqnarray}
\Delta{\bf P}_{ion} = \frac{\big | e \big | z_{\tau} \Delta u}{\Omega}
\end{eqnarray}

where $z_{\tau}$ is the valence atomic number of the $\tau^{th}$
atom. For practical purposes, the electronic contribution to
$Z^{*}$ can be obtained as

\begin{eqnarray}
Z^{el}_{\tau\alpha\beta} = \Omega \lim_{\Delta u_{\tau\alpha} \rightarrow 0} \frac{\Delta{\bf P}_{e}}{\Delta u_{\tau\alpha}}.
\end{eqnarray}
In periodic systems, the change in electronic polarization in zero field can be
computed from the King-Smith and Vanderbilt formula~\cite{kingsmith93}:

\begin{eqnarray}
{\bf P}^{el}_{\beta} = - \frac{1}{(2\pi)^{3}} i \Sigma_{n}^{occ} s\int_{BZ}
\langle u_{nk}|/\partial/\partial k_{\beta}|u_{nk}\rangle dk
\end{eqnarray}
 where $s$ is the occupation number of the valence band states ($s$ = 1 for spin-polarized case)
and $u_{nk}$ are the periodic parts of the Bloch functions. This
definition is valid only under the constrain that the wave
functions fulfil the periodic gauge condition. This means that the
periodic parts of the Bloch functions must satisfy

\begin{eqnarray}
u_{nk}(r) = e^{i\bf{G}.r} u_{nk+\bf{G}(r)}.
\end{eqnarray}

Once $\Delta{\bf P}$ is known, the Born effective charge tensor
for the $\beta$ component can be obtained from the formula
\ref{eqn:z} where $\alpha$ denotes the direction of polarization.
For a continuous change, such as the induced polarization when
atoms are displaced, uncertainties by possible integral multiples
of $2\pi$ can be avoided by defining small enough changes where
one always knows that the change in polarization is given by the
smallest value, i.e., with the integer equal zero. Hence, in the
Berry-phase calculations, we have chosen the displacement of
0.03{\AA}, and used a set of strings of 8-{\bf k} points (parallel
to some chosen reciprocal lattice vectors) to calculate the
electronic polarization. Our results shown in Table
\ref{table:bcharge} satisfy the acoustic sum rule,
$\sum_{k}Z_{k,ii}^{*}=0$ , indicating that the calculations are
well converged with respect to computational conditions.
\begin{table*}
\caption{Calculated Born effective charge tensor for
antiferromagnetic BiFeO$_3$ in the paraelectric
 $R$\={3}$c$ and ferroelectric $R3c$  structures} { \scriptsize
\begin{ruledtabular}
\begin{tabular}{l l c c c c c c c c c}
$Z^{*}$ & Position  & $xx$ & $yy$ & $zz$ & $xy$ & $xz$ & $yx$ & $yz$ & $zx$ & $zy$ \\
\hline
\\
$R$\={3}$c$\\
Bi & (1a) & 5.048 & 4.975 & 4.243 & 0.104 & $-$0.058 & $-$0.081 & $-$0.058 & $-$0.001 & 0.000 \\
Bi & (1a) & 5.048 & 5.071 & 4.240 & $-$0.105& $-$0.058 & 0.0330 & $-$0.058 & 0.000 & 0.000 \\
Fe & (1a) & 4.537 & 4.607 & 3.205 & $-$0.185& $-$0.069 & 0.2170 & $-$0.072 & 0.000 & 0.000 \\
Fe & (1a) & 4.537 & 4.568 & 3.207 & 0.186 & $-$0.068 & $-$0.115 & $-$0.067 & 0.000 & 0.000 \\
O  & (3b) & $-$3.799& $-$2.601& $-$2.583& 0.149 & 0.265  & $-$0.038 & 0.629  & 0.211 & 0.608 \\
O  & (3b) & $-$2.978& $-$3.469& $-$2.584& 0.418 & 0.434  & 0.636  & 0.554  & 0.421 & 0.486 \\
\\
$R3c$ \\
Bi  & (1a) & 5.438 & 5.460 & 4.792 & $-$0.022 & $-$0.004 & $-$0.001 & 0.004  & 0.000 & 0.000 \\
Bi  & (1a) & 5.434 & 5.414 & 4.791 & 0.022  & 0.005  & 0.001  & $-$0.005 & $-$0.001 & 0.000 \\
Fe  & (1a) & 5.224 & 5.223 & 4.859 & $-$0.777 & $-$0.001 & 0.777  & 0.000  & 0.000 & 0.000 \\
Fe  & (1a) & 5.224 & 5.222 & 4.858 & 0.777  & 0.000  & $-$0.777 & 0.000  & 0.000 & 0.000 \\
O   & (3b) & $-$3.590& $-$3.496& $-$3.227& $-$0.048 & 0.409  & 0.000  & 0.000  & 0.530 & $-$0.023\\
O   & (3b) & $-$3.551& $-$3.533& $-$3.226& $-$0.062 & $-$0.212 & $-$0.032 & $-$0.347 & $-$0.285& $-$0.446\\
\end{tabular}
\end{ruledtabular}
} \label{table:bcharge}
\end{table*}
\par
Our calculated $zz$ component of the BEC  can directly be compared
with the BEC values given in Table IV in
Ref.\,\onlinecite{neaton05}. The formal valence of Bi, Fe and O in
BiFeO$_3$ are 3+, 3+ and 2$-$ respectively. In simple,
high-symmetry oxides, the oxygen $Z^{*}$ is isotropic and is close
to $-$2~\cite{bilz79}. However, simple models~\cite{slater67} hint
at nontrivial values of $Z^{*}$. When we include AF ordering into
the calculations, because of the loss of symmetry, the space group
changed from $R3c$ to $R3$. Owing to the site symmetry, the
diagonal components of $Z^{*}$ are anisotropic for all the ions in
Table.\ref{table:bcharge}. If ions have closed-shell-like
character, each ion will carry an effective charge close to their
nominal ionic value according to a rigid-ion picture. On the
contrary, a large amount of nonrigid delocalized charge flows
across this compound during the displacement of ions owing to the
covalence effect~\cite{posternak94,zhong94}. Consequently, we have
obtained effective-charges much larger than the nominal ionic
value. The charges on Bi, Fe and O are around 75$\%$ larger than
they would have been in a pure ionic picture: this reveals the
presence of a large dynamic contribution superimposed to the
static charge. Similar giant values of $Z^{*}$ have been reported
using quite different technical ingredients and/or for other
perovskite oxides~\cite{zhong94,ghosez95}.

\par
The BEC is indeed a macroscopic concept~\cite{martin74,pick70},
involving the polarization of valence electrons as a whole, while
the charge ``belonging" to a given ion is an ill-defined concept.
The high BEC values indicate that relative displacements of
neighboring ions against each other trigger high polarization.
Roughly speaking, a large amount of nonrigid, delocalized charge
is responsible for higher value of BEC than the nominal charges.
The calculated BEC for the ferroelectric and the paraelectric
phases are given in Table\,\ref{table:bcharge}. It is clear that
both Bi and Fe donate electrons and O accepts electrons,
consistent with the traditional ionic picture. It can be recalled
that for a pure ionic system one could expect more isotropic value
for Born effective charges. Considerable anisotropy in the
diagonal components of BEC given in Table\,\ref{table:bcharge} and
the noticeable off-diagonal components in the oxygen sites confirm
the presence of covalent bonding between O 2$p$ and Fe 3$d$
orbitals. It is generally expected that there is considerable
covalent bonding between Fe and O within the FeO$_6$ octahedra and
this could explain the presence of anomalous contributions
(defined as the additional charge with respect to the well known
ionic value) to the effective charge at the Fe and O sites.
Surprisingly large anomalous charge at the Bi site (+1.8 to +2.4)
gives tangible proof for hybridization interaction between Bi and
the neighboring O atoms. When going from the paraelectric phase to
the ferroelectric phase, we have observed considerable increase in
the effective charges in all three constituents. This is
associated with the anisotropy in the chemical environment. While
FeO$_6$ octahedra are equally distanced from Bi ions in the
paraelectric phase, a short and a long Bi$-$FeO$_6$ distance arise
in the ferroelectric phase owing to the off-center displacement of
the Bi ions.


\subsection{Spontaneous polarization}

One of the basic quantities of ferroelectrics is spontaneous
macroscopic polarization {\bf P}, which results upon application
of an electric field, and persists at zero field in two (or more)
enantiomorphous metastable states of the crystal. Experiments
measure {\bf P} via hysteresis cycles from the difference $\Delta
{\bf P}$ between these metastable states~\cite{lines77}.  Even
though BiFeO$_3$ has large atomic displacements relative to the
centrosymmetric cubic perovskite structure, and the high
ferroelectric Curie temperature (1100\,K), early measurements
yielded rather small
polarizations~\cite{teague70,palkar02,wang04}. The small values
are in sharp contrast to the experimental findings on epitaxial
thin films~\cite{wang03,yun04,yun03} which were found to possess
large polarizations. There are several plausible explanations
given for the large difference in values of polarization reported
experimentally. One possibility is that the original reports of
small polarization might have been limited by poor sample
quality~\cite{teague70}. A second possibility is that the small
values could be correct for the $R3c$ structure, with the large
values being correct for different structural modifications
stabilized in the thin films~\cite{wang03}. Recently the large and
small values are explained within the modern theory of
polarization, which recognized that polarization is in fact a
lattice of values, rather than a single vector~\cite{neaton05}.
Hence it is interesting to analyze  the spontaneous polarization
present in BiFeO$_3$ in detail  with the help of accurate density
functional calculations.
\par
We calculated the polarization difference $\Delta {\bf P}$ between
the polar ($R3c;$ $\lambda$=1) and the ideal ($R$\={3}$c$;
$\lambda$=0) rhombohedral structures assuming a continuous
adiabatic transformation which consists of scaling the internal
strain with the parameter $\lambda$ ($0\leq\lambda\leq u)$. The
polarization at the paraelectric $R$\={3}$c$ phase [{\bf P(0)}]
would be zero (since there is no dipole for this case) and would
be a finite value {\bf P(u)} for the ferroelectric $R3c$ phase.
The dipole moment evolves continuously from {\bf P(0)}=0 to {\bf
P(u)}. i.e.

\begin{eqnarray}
\int_{0}^{u} Z^{*}(u)du = [{\bf P}(u)-{\bf P}(0)] = uZ(u)
\end{eqnarray}
where
\begin{eqnarray}
Z(u) = \frac{1}{u} \int_{0}^{u} Z^{*}(u)du
\end{eqnarray}

the mean value of $Z^{*}(u)$ from 0 to ${u}$ is equal to $Z(u)$.
We have calculated the polarization along  $x$, $y$ and $z$
directions using three different approximations in
Table\,\ref{table:pol}  i.e. By assuming (i.) the $Z(u)$ values
same as $Z^{*}(u)$ values of ferroelectric phase, (ii.)  the
$Z(u)$ values are same as $Z^{*}(u)$ values of paraelectric phase,
and (iii.) the $Z(u)$ values is an average of $Z^{*}(u)$ value of
the ferro-, and para-electric phases. The $Z_{zz}^{*}$ eigenvalues
of Bi and Fe correspond to an eigenvector aligned along the
ferroelectric axis. Hence, the calculated polarization values
corresponding to {\bf P$_{z}$} obtained from (iii) can be directly
compared with the experimentally-measured values.

\par
The $R3c$ symmetry permits the development of a spontaneous
polarization along [111], and Bi, Fe, and O are displaced relative
to one another along this threefold axis. In the case of O, the
eigenvector associated with the highest eigenvalue is positioned
approximately along the direction of Fe$-$O bond. We identify that
the highest contribution originates from the equatorial plane of
the FeO$_6$ octahedron. The largest relative displacements are
those of Bi relative to O owing to the presence of
stereochemically active Bi lone pair~\cite{seshadri011}. The
off-center displacements are noticeably larger compared to those
in nonlone-pair-active perovskite ferroelectrics such as BaTiO$_3$
or LiNbO$_3$~\cite{cohen92,inbar95}, but are consistent with those
observed in other Bi-based perovskites~\cite{iniguez03}.

\par
Our partial polarization analysis shows that the polarization
contribution coming from Bi, Fe, and O atoms are 86.94, 18.91, and
$-$17.15\,$\mu C$/cm$^{2}$ respectively along [111] direction.
This indicates that the polarization contribution arising from
displacement of Fe and O atoms are almost cancelling out and more
than 98~\% of the net polarization present in BiFeO$_3$ is
contributed by the Bi ions. The oxygen ions are displaced in the
same direction as Bi and Fe. However, the negative value of Born
effective charges at the oxygen sites is responsible for the
opposite direction of polarization at oxygen sites compared with
that in Bi and Fe sites. We have also calculated the polarization
using the point charge model and it is found to be around 32~\%
smaller than the actual value. The reduced value of polarization
obtained from point charge model is related to the covalence
effect.

\begin{table}
\caption { Cartesian components of the polarization $\Delta$ {\bf
P}$_{\alpha}$ for BiFeO$_3$ calculated using Eq.\ref{eqn:pol}
along a path from the paraelectric $R$\={3}$c$ phase to the
ferroelectric $R3c$ phase. The polarization values (in $\mu
C$/cm$^{2}$) calculated with the Born effective charges for
$R$\={3}$c$, $R3c$ structures and the average value for these two
structures (noted by ``average").}
\begin{ruledtabular}
\begin{tabular}{l c c c}
Z$^{*}$ values used in Eq.\,\ref{eqn:pol} & {\bf P}$_{x}$ & {\bf P}$_{y}$ & {\bf P}$_{z}$\\
\hline
\\
$R$\={3}$c$ & -0.02 & -0.01 & 93.85 \\
$R3c$ & 1.33 & 2.85 & 83.53 \\
Average & 0.65 & 1.41 & 88.69 \\
Point charge & 0 &  0 & 60.19
\end{tabular}
\end{ruledtabular}
\label{table:pol}
\end{table}

\par
From Table\,\ref{table:pol} it is clear that the polarization
values are very much directional dependent (see
Table~\ref{table:pol}). Our calculated value of polarization
corresponding to the $z$ direction can be compared with the
theoretical values of 84.2\,$\mu C$/cm$^{2}$ from
LSDA~\cite{neaton05} and 98.9\,$\mu C$/cm$^{2}$ from LDA+$U$
method~\cite{ederer05}. The large anisotropy in the calculated
polarization values indicate that if the samples are aligned along
[111], then they will have larger polarization than those aligned
along other orientations in the substrate. This could explain why
some reports show large value (35.7 $-$ 158\,$\mu C$/cm$^{2}$) of
polarization in this material and other show very small value (2.2
$-$ 8.9\,$\mu C$/cm$^{2}$). We found that the natural value of the
polarization is 88.7\,$\mu$C/cm$^{2}$ along [111] direction,
consistent with the recent measurements~\cite{wang03,yun04}.

\subsection{Summary}
\label{sec:con}

We have reported results from electronic structure calculations
for the multiferroic BiFeO$_3$ using generalized gradient
corrected accurate density functional calculation. We have found
that  A-AF-like phase is stable in energy compared to the other
magnetic configurations in BiFeO$_3$. This phase sustains
ferroelectric rhombohedral distortions with large shifts of Bi
atoms, in agreement with the fact that a large portion of the
ferroelectric polarization in Bi-based perovskite-like phases is
provided by Bi displacements. The stereochemically active lone
pair in BiFeO$_3$ causes a symmetry lowering structural distortion
which drives the ferroelectric phase transition and, as a
secondary effect, allows A-AF ordering or the canted spin
structure over the G-AF ordering. The agreement between the
calculated spontaneous polarization ({\bf P}) and recent
experimental data demonstrates that the crystal wave functions in
the frozen ferroelectric structure convey the essential physics of
the spontaneous polarization in real materials. The important
observations from our calculations can be summarized as follows.

\noindent
 1. Unlike earlier calculations, our calculated
structural parameters for ferroelectric BiFeO$_3$ are found to be
in good agreement with experimental values and especially the
experimentally-observed rhombohedral angle for this material has
been reproduced quantitatively.

\noindent 2. Structural stability studies show a pressure induced
structural transition from polar $R3c$ structure to the nonpolar
orthorhombic $Pnma$ structure around 130\,kbar. We have also
identified two ferroelectric phases such as tetragonal $P4mm$ and
monoclinic $Cm$ structures at the expanded lattice.

\noindent 3. We have observed insulating behavior in both the
paraelectric $R$\={3}$c$ phase and the ferroelectric $R3c$ phase
without including strong Coulomb correlation effect into the
calculations.

\noindent 4. Total energy calculations show large energy
difference between different magnetic configurations in BiFeO$_3$
and we found insulating behavior in A-AF and G-AF configuration
and metallic behavior in the nonmagnetic, ferromagnetic and C-AF
configurations. We believe that GGA itself sufficient to describe
the properties of BiFeO$_3$. However, if correlation effect is
significant in this material one could expect that the predicted
energetics of different magnetic configurations and electronic
behavior will change.

\noindent 5. Detailed analysis of the chemical bonding shows that
the electrons transfer from both Bi and Fe atoms to the  O sites
and also finite hybridization interaction takes place between the
Bi-O as well as Fe-O. This hybridization interaction plays a
dominant role in the large ferroelectric polarization.

\noindent 6. The role of Bi lone pair on the chemical bonding and
ferroelectric distortion has been analyzed. We found that the lone
pair lobe formation is driven primarily by covalency between Bi
$s$ and O $s/p$ states with the mediation of Bi $p$ states rather
than the classical picture of hybridization between Bi 6$s$ and Bi
6$p$ orbitals.

\noindent 7. The complete BEC tensor is reported for all the
constituents in both the ferroelectric and paraelectric phases
using accurate density functional calculations. The calculated BEC
at the ferroelectric and the paraelectric phase indicate that the
charges are strongly affected by ferroelectric displacement of
ions.

\noindent 8. Consistent with experimental observation, our
calculations shows that the easy axis of polarization is along
[111] direction in BiFeO$_3$. The ferroelectricity in BiFeO$_3$ is
originating from the distortion of Bi-O coordination environment
as a result of the stereochemical activity of the lone pair on Bi.

 \noindent 9. Large anisotropy in the polarization in
BiFeO$_3$ has been identified and this could explain why there is
large scattering in the experimentally reported polarization
values in this material.

 \noindent 10. Our partial polarization
analysis demonstrates that the polarization is essentially due to
the displacement of Bi relative to the center of the FeO$_6$
octahedra.

\par
The overall conclusion is that our results are in good agreement
with experiment. This demonstrates that  accurate density
functional calculations can be used to explore not only the
structural and magnetic properties but also the ferroelectric
properties of magnetoelectric materials.

 \acknowledgements
 The authors gratefully acknowledge the Research Council of Norway for
financial support under the grant number 158518/431 (NANOMAT) and
for the computer time at the Norwegian supercomputer facilities.
P.R. wishes to thank useful communications received from Florent
Boucher, Jan Lazewski, Jeffrey Neaton, Karen Johnston, Umesh
V.Waghmare, Sergey Prosandeev, Arne Klaveness, and Karel Knizek.
The skillful assistance from the staff at Swiss-Norwegian Beam
Line, ESRF is gratefully acknowledged.

\end{document}